\documentclass[preprint2]{proto}
\usepackage{times} 

\newcommand{\refs}{\par\noindent\hangindent=1pc\hangafter=1}
\begin{document}

\title{\textbf{\LARGE The Taurus Molecular Cloud: Multi-Wavelength Surveys with XMM-Newton,
    the Spitzer Space Telescope, and CFHT}\footnote{With significant contributions from: Lori Allen (CfA),
         Marc Audard (Columbia University),
         Jer\^ome Bouvier (Grenoble), Kevin Briggs (PSI), Sean Carey (Caltech), Elena Franciosini (Palermo), Misato Fukagawa (Caltech), 
	 Nicolas Grosso (Grenoble), Sylvain Guieu (Grenoble), Dean Hines (Space Science Institute), Tracy Huard (CfA), 
	 Eugene Magnier (IfA Honolulu), Eduardo Mart\'{\i}n (IAC, Spain), Fran\c{c}ois M\'enard (Grenoble), Jean-Louis Monin (Grenoble), Alberto Noriega-Crespo (Caltech),
	 Francesco Palla (Florence), Luisa Rebull (Caltech),
	 Luigi Scelsi (Palermo), Alessandra Telleschi (PSI), Susan Terebey (CalStateLA).
	 }}
 
\author {\textbf{\large Manuel G\"udel}}
\affil{\small\em Paul Scherrer Institut} 

\author {\textbf{\large Deborah L. Padgett}}
\affil{\small\em California Institute of Technology} 

\author {\textbf{\large Catherine Dougados}}
\affil{\small\em Laboratoire d'Astrophysique de Grenoble}

\begin{abstract}
\baselineskip = 11pt
\leftskip = 0.65in 
\rightskip = 0.65in
\parindent=1pc
{\small
The Taurus Molecular Cloud (TMC) ranks among the nearest and
best-studied low-mass star formation regions. It contains numerous
prototypical examples of deeply embedded protostars with massive
disks and outflows, classical and weak-lined T Tauri stars, jets
and Herbig-Haro objects, and a growing number of confirmed brown
dwarfs. Star formation is ongoing, and the cloud covers all stages
of pre-main sequence stellar evolution. We have initiated
comprehensive surveys of the TMC, in particular including:
(i)   a deep X-ray survey of about 5 sq. degrees with
      {\it XMM-Newton};
(ii)  a near-to-mid-infrared photometric survey of $\approx$30 sq. degrees
      with the Spitzer Space Telescope, mapping the entire cloud
      in all available photometric bands; and
(iii) a deep optical  survey using the
      Canada-France-Hawaii Telescope.
Each wavelength regime contributes to the understanding of
different aspects of young stellar systems. {\it XMM-Newton} and Spitzer
mapping of the central TMC is a real breakthrough in disk
characterization,  offering the most detailed studies of correlations
between disk properties and high-energy magnetic processes in any
low-mass star-forming region, extending also to brown dwarfs in
which disk physics is largely unexplored. The optical data
critically complements the other two surveys by allowing clear source
identification with 0.8$^{\prime\prime}$ resolution, identifying substellar candidates,
and, when combined with NIR data, providing the wavelength baseline
to probe NIR excess emission. We  report results and correlation
studies from these surveys. In particular, we address the physical
interpretation of our new X-ray data, discuss the entire young stellar
population from embedded protostars to weak-lined T Tau stars and
their environment, and present new results on the low-mass population
of the TMC, including young brown dwarfs.
 \\~\\~\\~}
\end{abstract}

\section{\textbf{INTRODUCTION}}\label{intro}

In a modern picture of star formation, complex feedback loops regulate mass accretion processes, 
the ejection of jets and outflows, and the chemical and physical evolution of disk material destined to
form planets. Observations in X-rays  with {\it Chandra} and {\it {\it XMM-Newton}} penetrate dense molecular 
envelopes, revealing an environment exposed to high levels of X-ray radiation. In a complementary 
manner, observations in the infrared with the {\it Spitzer Space Telescope} ({\it Spitzer}) now obtain  
detailed infrared photometry and spectroscopy with diagnostics for disk structure and chemical 
composition of the gas  and dust in the circumstellar environment. Furthermore, optical surveys have reached
a sensitivity and area coverage with which a detailed census of the substellar population has become
possible.

Near- to far-infrared (IR) emission originates predominantly in the dusty environment of the forming
stars, either in contracting gaseous envelopes or in circumstellar disks. IR excess (relative
to the photospheric contributions) has been successfully used to model disk geometry, the
structure of the envelope, and also the composition and structure of dust grains 
({\it d'Alessio et al.,} 1999).

X-rays play a crucial role in studies of star formation, both physically and diagnostically.
They may be generated at various locations in young stellar systems, such as in a ``solar-like''
coronal/magnetospheric environment, in shocks forming in accretion funnel flows (e.g., {\it Kastner et 
al.,} 2002), or in jets and Herbig-Haro flows (e.g., {\it Pravdo et al.,} 2001; {\it Bally 
et al.,} 2003; {\it G\"udel et al.,} 2005). 
By ionizing circumstellar material, X-rays also determine some of the prevalent 
chemistry ({\it Glassgold et al.,} 2004) while making the gas accessible to magnetic fields. 
Ionization of the circumstellar-disk surface may further drive 
accretion instabilities ({\it Balbus and Hawley,} 1991). Many of these X-ray related
issues are summarized in the chapter by {\it Feigelson et al.} in this volume, based on recent 
X-ray observations of star-forming regions.

\bigskip
\noindent
\textbf{ 1.1 The Taurus Molecular Cloud Complex}
\bigskip

The Taurus Molecular Cloud (TMC henceforth) has played a fundamental role in our understanding
of low-mass star formation. At a distance around 140~pc (e.g., {\it Loinard et al.,} 2005;
{\it Kenyon et al.,} 1994), it is  one of the nearest star formation
regions (SFR) and reveals characteristics that make it ideal for detailed physical studies.
One of the most notable properties of the TMC in this regard is its structure in which several loosely
associated but otherwise rather isolated molecular cores each produce one or only a few low-mass
stars, different from the much denser cores in $\rho$ Oph or in Orion. TMC features a low 
stellar density of only 
1--10 stars~pc$^{-2}$ (e.g., {\it G\'omez et al.,} 1993). Strong mutual influence due
to outflows, jets, or gravitational effects are therefore minimized. Further, most stars in
TMC are subject to relatively modest extinction, providing access to a broad spectrum of 
stars at all evolutionary stages from Class 0 sources to near-zero age main-sequence 
T Tau stars. TMC has also become of central interest for the study of substellar
objects, in particular brown dwarfs (BD), with regard to their evolutionary history and their
spatial distribution and dispersal ({\it Reipurth and Clarke,} 2001; {\it Brice\~no et al.,} 2002).

TMC has figured prominently in star-formation studies at all wavelengths. It has provided
the best-characterized sample of classical and weak-lined T Tau stars (CTTS and WTTS, respectively, 
or ``Class II'' and ``Class III'' objects - {\it Kenyon and Hartmann,} 1995); most of our current picture
of low-density star formation is indeed based on IRAS studies of TMC ({\it Beichman et al.,} 1986; 
{\it Myers et al.,} 1987; {\it Strom et al.,} 1989; {\it Kenyon et al.,} 1990; {\it Weaver and Jones,} 1992;
and {\it Kenyon and Hartmann,} 1995).  Among the key results from TMC studies as listed in {\it Kenyon and Hartmann} (1995) 
figure the following: i) More than 50\% of the TMC objects have IR excess beyond the photospheric
contribution, correlating with other activity indicators (H$\alpha$, UV excess etc.) and indicating
the presence of warm circumstellar material predominantly in the form of envelopes for Class I
protostars and circumstellar disks for Class II stars. ii) Class III sources (mostly to be 
identified with WTTS) are distinctly different from lower classes by not revealing optically thick disks or 
signatures of accretion. iii) Star formation has been ongoing at a similar level during the past
1-2~Myr, with the Class-I protostars having ages of typically 0.1-0.2~Myr. iv) There is clear support
for an evolutionary sequence Class I$\rightarrow$II$\rightarrow$III, although there is little 
luminosity evolution along with this sequence, indicating different evolutionary
speeds for different objects. The infall time scale is a few times $10^5$~yrs, while the disk phase 
amounts to a few times $10^6$~yrs.

TMC has also been well-studied at millimeter wavelengths, having better
high-resolution molecular line maps than any other star-forming region ({\it Onishi et al.} 2002).
This region has been a major site for searches of complex molecules ({\it Kaifu et al.,} 2004); many
molecular transitions have been mapped across a large area, and the results will be interesting to compare with
large-scale IR dust maps of the TMC (see Fig.~\ref{ff3} below).   

In X-rays, TMC has again played a key role in our understanding of high-energy processes
and circumstellar magnetic fields around pre-main sequence stars. Among the key surveys are those 
by {\it Feigelson et al.} (1987), {\it Walter et al.} (1988), {\it Bouvier} (1990), {\it Strom et al.} (1990), 
{\it Damiani et al.} (1995) and {\it Damiani and Micela} (1995)  based on {\it Einstein Observatory} observations,
and the work by {\it Strom and Strom} (1994), {\it Neuh\"auser et al.} (1995) and {\it Stelzer and Neuh\"auser} (2001)
based on {\it ROSAT}. 
These surveys have characterized the overall luminosity behavior of TTS, indicated a dependence of
X-ray activity on rotation, and partly suggested X-ray differences between CTTS and WTTS. 

\bigskip
\noindent
\textbf{ 1.2 Is TMC Anomalous?}
\bigskip

Although TMC has been regarded, together with the $\rho$ Oph dark cloud, as the prototypical
low-mass star-forming region, a few apparent peculiarities should be mentioned. TMC contains
an anomalously large fraction of binaries ({\it Ghez et al.,} 1993), 
compared with other SFRs  (e.g., Orion) or with field stars.
In TMC, about two thirds of all members are bound in multiple systems, with an average
separation of about $0.3^{\prime\prime}$ (e.g., {\it Leinert et al.,} 1993; {\it Mathieu,} 1994; 
{\it Simon et al.,} 1995; {\it Duch\^ene et al.,} 1999; {\it White and Ghez,} 2001; {\it Hartigan and Kenyon,} 
2003). Also, TMC cloud cores are comparatively small and low-mass 
when compared with cores in Orion or Perseus ({\it Kun} 1998). 

TMC has also been found to be deficient of lowest-mass stars and BDs, with a mass 
distribution significantly enriched in 0.5--1~M$_{\odot}$ stars, compared to Orion samples 
({\it Brice\~no et al.,} 2002). 
The formation of BDs may simply be different in the low-density environment of TMC
compared to the dense packing of stars in Orion.

Further, for reasons very poorly understood, TMC differs from other SFRs
significantly with regard to X-ray properties. Whereas no X-ray activity-rotation correlation 
(analogous to that in main-sequence stars) is found for samples in the Orion star-forming regions,
perhaps suggesting that all stars are in a saturated 
state ({\it Flaccomio et al.,} 2003a; {\it Preibisch et al.}, 2005a), the X-ray activity  
in TMC stars has been reported to decrease for increasing rotation period 
(e.g., {\it Neuh\"auser et al.,} 1995; {\it Damiani and Micela,} 1995; {\it Stelzer and 
Neuh\"auser,} 2001). Also, claims have been made
that the X-ray behavior of TMC CTTS and WTTS 
is significantly different, CTTS being less luminous than WTTS ({\it Strom and Strom,} 1994; 
{\it Damiani et al.,} 1995; {\it Neuh\"auser et al.,} 1995; {\it Stelzer and Neuh\"auser,} 2001). This  
contrasts with other star-forming regions ({\it Flaccomio et al.,} 2000; {\it Preibisch and 
Zinnecker,} 2001), although recent reports reveal a similar 
segregation also for Orion and some other SFRs ({\it Flaccomio et al.,} 2003b; {\it Preibisch et 
al.,} 2005a). Some of these discrepancies may be due to selection and detection bias 
(e.g., WTTS are predominantly identified in X-ray studies, in contrast to CTTS).

\bigskip
\centerline{\textbf{ 2. NEW SURVEYS OF THE TAURUS CLOUDS}}
\bigskip

\noindent
\textbf{ 2.1 The Need for New Surveys}
\bigskip

Some of the anomalies mentioned above may be strongly influenced
by survey bias from  the large size of TMC (25+ sq. degrees). For example, with the exception
of IRAS and 2MASS, all the IR surveys of TMC focus on limited regions around dense 
molecular cloud cores of TMC (or of other star-forming regions) since these are easy to map 
using inefficient point-and-shoot strategies with small IR arrays. Systematic and deep 
investigations of more distributed low-mass star formation such as that in TMC are lacking.  
The exception in X-rays is the ROSAT All-Sky Survey obtained in the early nineties. However, 
this survey was of comparatively low sensitivity and recorded only rather soft X-ray photons, 
thus leaving all embedded sources and most of the lowest-mass stellar and BD population 
undetected. Angular resolution has also provided serious ambiguity.

A suite of new telescopes and satellite observatories now permits to reconsider the issues
mentioned in Section~\ref{intro}.
We have started a large multi-wavelength project to map significant portions of TMC
in X-rays, the optical,  near-infrared, and mid-infrared. 
The instruments used, {\it XMM-Newton} in X-rays, {\it Spitzer} in the near-/mid-infrared range, and the 
France-Canada-Hawaii Telescope (CFHT) in the optical range, provide an observational
leap forward. Table~\ref{instruments} compares characteristic limitations of our surveys
with limitations defined by previous instruments, represented here by the ROSAT 
Position-Sensitive Proportional Counter (PSPC All-Sky Survey for X-rays, {\it Neuh\"auser et al.,} 1995; also some
pointing observations, {\it Stelzer and Neuh\"auser,} 2001), IRAS (25\micron\ for mid-infrared data, 
{\it Kenyon and Hartmann,} 1995), and the KPNO 0.9~m telescope CCD surveys (for the optical, 
{\it Brice\~no et al.,} 1998, 2002; {\it Luhman} 2000; {\it Luhman et al.} 2003; {\it Luhman} 2004).

\begin{table}[h]
\small
\caption{Characteristics of  previous and new  TMC surveys}
\label{instruments}
\begin{tabular}{lll} 
\hline
\hline
 Parameter                                   &  Previous surveys                    & New surveys              \\
\hline
\multicolumn{3}{l}{Luminosity detection limit:}                                         \\
\quad X-rays [erg~s$^{-1}$]                  & $\approx 10^{29}$            & $\approx 10^{28}$            \\
\quad mid-IR [$L_{\odot}$]                   & $>0.1$                       & 0.001                     \\
\quad optical$^a$ [$L_{\odot}$]              & $\approx 10^{-3}$            & $\approx 10^{-4}$                       \\
\multicolumn{3}{l}{Flux detection limit:}                                          \\
\quad X-rays [erg~cm$^{-2}$s$^{-1}$]         & $\approx 4\times 10^{-14}$   & $\approx 4\times 10^{-15}$        \\
\quad mid-IR [$24\mu$m, mJy]                 & 500                          & 1.2                     \\
\quad optical [I, mag]                       & $\approx 21$                 & $\approx 24$                       \\
\multicolumn{3}{l}{Mass detection limit:}                                                      \\
\quad X-rays                                 & $\approx 0.1M_{\odot}$       & $\approx 0.05M_{\odot}$	     \\
\quad mid-IR, $24\mu$m                       & $\approx 0.5M_{\odot}$       & $\approx 1~M_{\rm Jup}$	    \\
\quad optical$^a$                            & $\approx 15-20~M_{\rm Jup} $ & $\approx 10~M_{\rm Jup} $       \\
\multicolumn{3}{l}{Accessible object types:$^b$}                                   \\
\quad X-rays                                 & II,III		            & I-III, BD, (HH)  \\
\quad mid-IR                                 & 0-III            	    & 0-III, BD, Ph, DD	  \\
\quad optical                                & I-III, BD                    & I-III, BD   	  \\
\multicolumn{3}{l}{Angular resolution (FWHM, resp. pixel size):}                                                \\
\quad X-rays                                 & $\approx  25^{\prime\prime}$ & $\approx 4.5^{\prime\prime}$     \\
\quad mid-IR, 24$\mu$m                       &$1^{\prime}\times 5^{\prime}$ & $\approx 6^{\prime\prime}$     \\
\quad optical                                & $\approx 0.67^{\prime\prime}$/pix   & $\approx 0.2^{\prime\prime}$/pix     \\
\multicolumn{3}{l}{Energy or wavelength range:}                                   \\
\quad X-rays [keV]                           & 0.1-2.4                      &  0.2-15     \\
\quad mid-IR [$\mu$m]                        & 1.2-100   	            &  3.6-160        \\
\quad optical [nm]                           & 800-1000		            &  780-1170        \\
\multicolumn{3}{l}{Spectral resolution or filter width:}                                                 \\
\quad X-rays [$E/\Delta E$]                  & $\approx  1$		    & $\approx$10-50$^c$    \\
\quad mid-IR $24\mu$m  [$\Delta\lambda$]     & $5.8\mu$m		    & $4.7\mu$m          \\
\multicolumn{3}{l}{Spatial extent:}                                                 \\
\quad X-rays                                 & all-sky		            & 5~sq. deg        \\
\quad mid-IR                                 & all-sky   		    & 30~sq. deg        \\
\quad optical                                & 12.4~sq. deg   		    & 34~sq. deg        \\
\hline
\multicolumn{3}{l}{$^a$ derived from the DUSTY models of {\it Chabrier et al.}}\\
\multicolumn{3}{l}{(2000) for $A_V < 5$~mag and age $< 5$~Myr}\\
\multicolumn{3}{l}{$^b$ IR classes; HH = Herbig-Haro, DD = debris disks,}\\
\multicolumn{3}{l}{Ph = photospheres (Class III)}\\
\multicolumn{3}{l}{$^c$ X-ray gratings: $\approx 300$ (selected targets)}\\
\normalsize
\end{tabular}
\vskip -0.7truecm\end{table}

\bigskip
\noindent
\textbf{ 2.2 Scientific Goals of the New Surveys}
\bigskip

\begin{figure*}[t!]
\centerline{\resizebox{0.7\hsize}{!}{\includegraphics{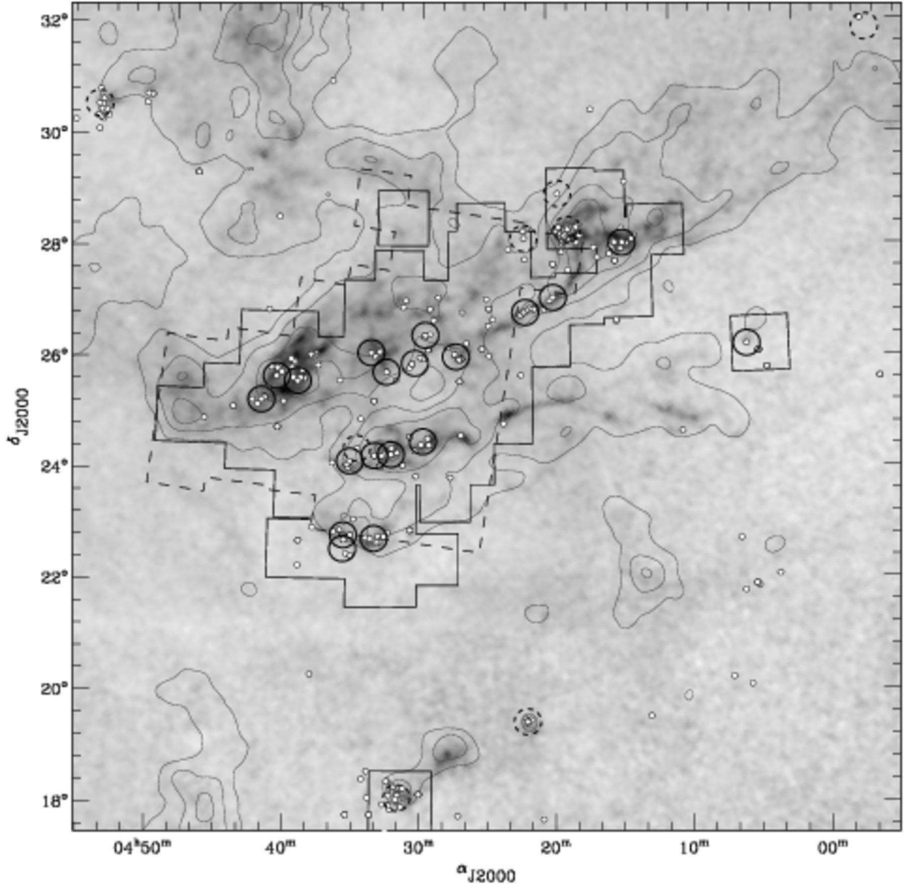}}}
\caption{\small Map of the TMC region. The grayscale background map is an extinction ($A_V$) map
from {\it Dobashi et al.} (2005). The outlines of the CFHT and the {\it Spitzer} surveys are indicated by the solid and
dashed polygons, respectively. The large (0.5 degrees diameter) circles show the fields of
view of the {\it XMM-Newton} X-ray survey (the dashed circles marking fields from separate projects also
used for the survey). Small white dots mark the positions of individual young stars. Note the outlying
{\it XMM-Newton} fields around SU Aur (NE corner), $\zeta$ Per (NW corner), and L1551 and T Tau (south).
(Figure courtesy of Nicolas Grosso.)\label{ff1}} 
\end{figure*}

\noindent {\it 2.2.1 The XMM-Newton X-Ray Survey.}
Our {\it XMM-Newton} X-ray survey
maps approximately 5 sq. degrees (Fig.~\ref{ff1}) of the denser molecular cloud areas with limiting 
sensitivities  around $L_X \approx 10^{28}$~erg~s$^{-1}$, sufficient, for the first time, 
to detect nearly every lightly absorbed, normal CTTS and WTTS  and a significant 
fraction of BDs and protostars. The goals of this survey are:

\noindent{1. To collect X-ray spectra and light curves from a statistically meaningful
           sample of TMC objects, and to characterize them in terms
	   of X-ray emission measure distributions,  temperatures, X-ray luminosities,
	   and variability.}

\noindent{2.  To interpret X-ray emission in the context of other stellar properties
           such as rotation, mass accretion and outflows.}

\noindent{3.  To investigate changes in the X-ray behavior as a young stellar object evolves.}

\noindent{4. To obtain a census of X-ray emitting objects at the stellar mass limit and
          in the substellar regime (BDs).}

\noindent{5. To study in what sense the stellar  environment influences the X-ray 
            production, and vice versa.}

\noindent{6. To search for new, hitherto unrecognized TMC members.}
\par
The outstanding characteristics of the survey are its sensitivity, its energy resolution, and
its energy coverage. The TMC population is detected nearly completely in the 
surveyed fields, thus suppressing potential bias that previous surveys may 
have been subject to.  In particular, a significant fraction of the TMC BDs can now
be studied and their high-energy properties  put into  context with T Tauri stars. 
Moderate energy resolution  permits a detailed description 
of the thermal plasma properties together with the measurement of the 
absorbing gas columns that are located predominantly in the Taurus clouds themselves, and even 
in the immediate circumstellar environment in the case of strongly absorbed objects.
Photoelectric absorption acts more strongly on the softer photons. ROSAT detected no
TMC protostars because photons below 2~keV are almost completely absorbed.
In contrast, harder photons to which {\it XMM-Newton} is sensitive 
penetrate large absorbing gas columns around protostars.

X-rays can be used to detect
new candidate members of the TMC. We note in particular that WTTS are best identified
through X-ray surveys ({\it Neuh\"auser et al.,} 1995). Placement on the HR diagram, X-ray spectral and
temporal properties, and infrared photometry together  can characterize new sources as
being likely members of the association.

\bigskip
\noindent {\it 2.2.2 The {\it Spitzer} Infrared Survey.}
We have mapped the majority of the main TMC
($\approx$30 sq. degrees, Fig.~\ref{ff1}) using the {\it Spitzer} IRAC and MIPS 
instruments in order to take a deep census of
young stars and disks to below the deuterium burning limit.
These data play a crucial role in characterizing circumstellar material and BDs in our 
multiwavelength surveys of TMC. The goals are:

\noindent{1. To survey the nearest example of a low-density star-forming region
completely and objectively for stars and BDs.}

\noindent{2. To learn whether there is a ``distributed'' component of isolated 
      star formation far removed from the multiple ``centers'' within the TMC complex.}

\noindent{3. To determine (via combination with ground-based data) what the 
distribution of stellar ages, masses and disk lifetimes are.}

\noindent{4. To carry out a definitive search for disks in transition 
between optically thick accretion disks and post-planet building disks 
and to use these data to constrain the timescales for planet building.}

\noindent{5. To discover which optical/2MASS/X-ray sources are the counterparts
of known IRAS sources, as well as new {\it Spitzer} far-IR sources.}

\bigskip
\noindent {\it 2.2.3 The CFHT Optical Survey.}
Our CFHT TMC survey was conducted as part of a larger-scale program
targeted towards young galactic clusters (PI: J. Bouvier). 
 It is a comprehensive, deep optical survey of the TMC
down to 24 mag in I (23 in z) of 34 sq. degrees, covering most of
the {\it XMM-Newton} and {\it Spitzer} fields (Fig.~\ref{ff1}). The specific goals
of the CFHT survey are: 

\noindent{1.  To obtain a complete census of the low-mass star population down to 
   I $\simeq$ 22, well into the substellar regime.} 

\noindent{2.  To obtain accurate optical I band photometry critical to derive 
fundamental parameters (reddening, luminosity estimates) for low-luminosity TMC members 
and investigate the occurrence of IR excesses.}

\noindent{3. To obtain deep images at sub-arcsecond resolution: 1) for a proper 
source identification, 2) to detect and map new large-scale HH flows and embedded envelopes. }

\noindent These characteristics make the TMC CFHT survey the largest optical survey of the 
TMC sensitive well into the substellar regime.  These data are critical for a proper
identification of the sources and the determination of their
fundamental parameters (effective temperature, reddening,
luminosities). In particular, they play a central role in the
detection and characterization of the TMC substellar population.

\bigskip
\centerline{\textbf{ 3. SURVEY STRATEGIES AND STATUS}}
\bigskip

\noindent
\textbf{3.1 The {\it XMM-Newton} Survey}
\bigskip

Our {\it XMM-Newton} survey  comprises 19 exposures obtained coherently
as part of a large project ({\it G\"udel et al.,} 2006a), complemented by an additional 
9 exposures obtained in separate projects, bringing the total exposure time to 1.3~Ms. 
The former set of exposures used all three EPIC
cameras in full-frame mode with the medium filter, collecting photons during
approximately 30~ks each. Most of the additional fields obtained
longer exposures (up to 120~ks) but the instrument set-up was otherwise similar. 
The spatial coverage of our survey is illustrated in Fig.~\ref{ff1}.
The survey includes approximately 5 sq. degrees in total, covering  a useful energy range of 
$\approx 0.3-10$~keV with a time resolution down to a few seconds. The circular
fields of view with a diameter of 0.5 degrees each
were selected such that they cover the densest concentrations of CO gas, which also
show the strongest accumulations of TMC stellar and substellar members.  We also obtained
near-ultraviolet observations with the Optical Monitor (OM), usually through
the U band filter, and in exceptional cases through filters at somewhat shorter
wavelengths. The OM field of view is rectangular, with a side length of $17^{\prime}$.  
An on-axis object could be observed at  high time resolution (0.5~seconds).

Data were processed using standard software and analysis tools.
Source identification was based on procedures involving maximum-likelihood 
and wavelet algorithms.  In order to optimize detection of faint sources,
periods of high background local particle radiation levels  were cut out. 

A typical exposure with an average background contamination level reached a detection
threshold of $\approx 9\times 10^{27}$~erg~s$^{-1}$ on-axis and  $\approx 1.3\times 
10^{28}$~erg~s$^{-1}$ at 10$^{\prime}$   off-axis for a lightly absorbed X-ray 
source with a thermal spectrum characteristic of T Tau stars (see below) subject to
a hydrogen  absorption column density of $N_H = 3\times 10^{21}$~cm$^{-2}$. This 
threshold turns out to be appropriate to detect nearly every
T Tau star  in the surveyed fields.

We interpreted all spectra based on thermal components combined in such a way
that they describe emission measure distributions similar to those previously 
found for nearby pre-main sequence stars or active zero-age main sequence 
stars ({\it Telleschi et al.,} 2005; {\it Argiroffi et al.,} 2004; {\it Garcia et al.,} 2005; 
{\it Scelsi et al.,} 2005). An absorbing hydrogen column density was also fitted to the
spectrum.

\bigskip
\noindent
\textbf{3.2 The {\it Spitzer} Survey}
\bigskip

In order to study material around very low-mass young stars
and BDs, we require the use of both the {\it Spitzer}
imaging instruments (IRAC and MIPS). 
The short IRAC bands detect the photospheres of BDs and
low mass stars. Using the models of {\it Burrows et al.} (2003), we find that 2--12~s 
exposures are required to detect BDs down to 1 M$_{\rm Jup}$
in the 4.5~\micron\ IRAC band (see Table~\ref{spitzersens}).
Young BDs without disks will require
spectroscopic confirmation since they will resemble M stars. The exquisite
sensitivity of IRAC also makes it susceptible to saturation issues
for the solar-type population of the Taurus clouds. In order to
mitigate these effects, we have observed with the ``high dynamic
range" mode which obtains a short (0.4~s) frame together with
a 10.4~s one. Total IRAC and MIPS sensitivities for our maps are listed
in Table~\ref{spitzersens}, together with the number of independent exposures
performed per point in the map (data redundancy).
Disks are indicated by measured flux densities well above photospheric 
levels and/or stellar colors inconsistent with a Rayleigh-Jeans 
spectrum between any pair of bands.
The presence of disks is revealed by comparison of the 8 and 24~\micron\ 
bands to each other and the shorter IRAC bands.
For MIPS, we chose fast scan for efficient mapping at 24~\micron.
The resulting 1.2~mJy 5$\sigma$ sensitivity for MIPS 24~\micron\  
enables detection of disks 20$\times$ fainter than the low-luminosity
edge-on disk HH 30 IRS ({\it Stapelfeldt et al.,} 1999). This is sufficient
to detect $\geq$ few M$_{\rm Jup}$ disks around $\geq$ 0.007~M$_{\odot}$ 
3~Myr old BDs ({\it Evans et al.,} 2003).

Due to the proximity of TMC, it subtends
an area of more than 25 sq. deg on the sky.
Spitzer is the first space observatory with modern sensitivity
which possesses the observing efficiency to map the region as a
single unit within a feasible observing time.
Our observing program attempted to survey the TMC 
nearly completely and objectively within a limited time
(134~hours). 
In particular, we chose our coverage of TMC (illustrated in
Fig.~\ref{ff1}) to overlap as many CFHT and {\it XMM-Newton} pointings 
as possible while mapping all the known contiguous $^{13}$CO cloud.
In addition, in order to obtain imaging of the
Heiles 2 and Lynds 1536 clouds most efficiently by mapping
in long strips, we obtained a considerable portion of 
adjacent off-cloud area south of the TMC. This region is
invaluable for assessing the distribution of sources
with excess off the cloud, as well as galactic and
extragalactic contamination. The MIPS data were obtained
using the scan mapping mode ({\it Rieke et al.,} 2004) which
uses a cryogenic mirror to perform image motion
compensation for short exposures while the telescope is
slewing. The ``fast scan" mode is by far the most
efficient way for {\it Spitzer} to map large areas
quickly, covering more than 1~sq. degree per hour.
Due to the
lack of an internal scan mirror, IRAC maps at
the slow speed of 0.33~sq. degrees per hour
at the minimal depth of our survey. Because
each point in the IRAC map has a
total redundancy of only two exposures, cosmic
ray removal is difficult. 

Since TMC is located virtually on the ecliptic plane, we mapped
twice at 8 and 24~\micron\ to identify asteroids. 
A waiting period of 3--6~hr
between mapping epochs allowed asteroids with a minimum
motion of 3.5\arcsec/hour to move enough so that they can be 
identified by comparing images from the two epochs. 
Initial results
suggest there are several thousand asteroids detected
in our maps. These will be eliminated from the catalog of 
young stellar objects. However, our strategy enables 
a statistical asteroid study with the largest low ecliptic latitude survey 
yet envisioned for {\it Spitzer}.

\begin{table}[h]
\caption{{\it Spitzer} TMC survey estimated sensitivity}
\label{spitzersens}
\begin{tabular}{cccc} 
\hline
\hline
 Instrument  &  Band             &  5$\sigma$ Point Source       & Redundancy \\
             &  (\micron)        &   Sensitivity (mJy)           &   \\
\hline
MIPS & 24 & 1.2 & 10 \\
MIPS & 70 & $\sim$44 & 5 \\
MIPS & 160 & $\sim$500; 2.64 MJy/sr & $\sim$1 \\
\nodata & \nodata & (extended source) & (bonus band) \\
IRAC & 3.6 & 0.017 & 2 \\
IRAC & 4.5 & 0.025 & 2 \\
IRAC & 5.8 & 0.155 & 2 \\
IRAC & 8.0 & 0.184 & 2 \\
\hline
\end{tabular}
\end{table}

\begin{figure}[h!]
\epsscale{0.5}
\centerline{\resizebox{0.85\hsize}{!}{\includegraphics{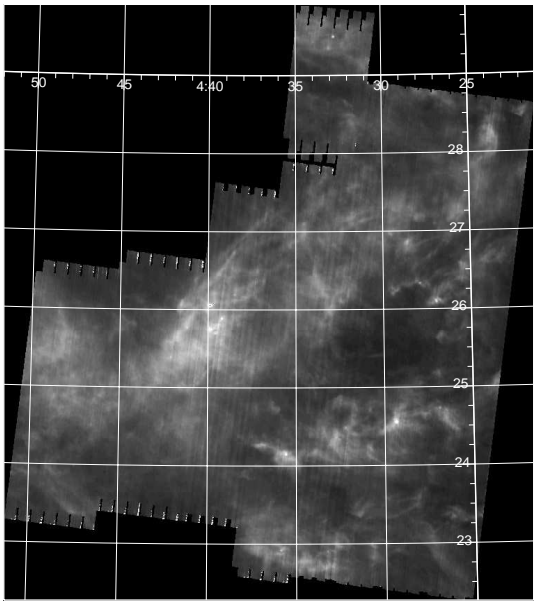}}}
\vskip -0.3truecm
\caption{\baselineskip  13pt \small 29 square degree MIPS 160~\micron\ map of nearly the entire
TMC region observed in the {\it Spitzer} TMC survey. The brightest regions
reach an intensity of over 200 MJy/sr. The image has been slightly smoothed
to fill in some narrow gaps in coverage. Effective exposure time is about
3 seconds. \label{ff2}}
\end{figure}

{\it Spitzer} observations of TMC were performed in February and March
of 2005. Although the bulk of the data were released to the
team in May, the field which includes Lynds 1495 remains embargoed
until March 2006. The team has spent several months in 2005 developing algorithms
to mitigate remaining instrumental signatures in the data,
many of which are aggravated by the presence of bright point sources
and extended nebulosity. Removal of radiation artifacts from the IRAC
data has been a considerable challenge due to the minimal repeat
coverage of our mapping strategy. However, as of PPV, a prototype
reprocessing code has been implemented for IRAC. Using this software
and a robust mosaicing algorithm which uses the short HDR frame to
eliminate the brightest cosmic rays, a 1.5 square degree region centered
around 04$^{\rm h}$~26$^{\rm m}$~10.9$^{\rm s}$~ 
+27$^{\circ}$~18$^{\prime}$~10$^{\prime\prime}$ (J2000)  
has been fully reprocessed in the IRAC
and MIPS bands, source extracted, and bandmerged.  
The MIPS 24~\micron\ data are most affected by the presence of thousands of
asteroids, which outnumber the stars at this wavelength.  
It is therefore necessary to separately extract
sources from each of the MIPS 24~\micron\ observing epochs, then
bandmerge these lists to achieve a reliable source list minus asteroids. 
One of the most interesting preliminary data
products is a complete 160~\micron\ map of the entire available TMC
region (29 square degrees). This map is presented in Fig.~\ref{ff2}.
The areas of brightest 160~\micron\ emission correspond well to
the $^{13}$CO clouds.

\bigskip
\noindent
\textbf{3.3 The {\it CFHT} Survey}
\bigskip

The CFHT TMC survey has been performed in 4 successive periods on
the Canada-France-Hawaii Telescope with the CFH12k and MEGACAM
large-scale optical cameras (see Table~\ref{tab:phot-survey} for a
detailed journal). Taking into account overlapping fields, a total
effective surface of 34~sq. degrees has been surveyed down to an I band sensitivity 
limit of I$_{C}$ = 24, i.e., well into the substellar regime (see below). 
It encompasses more than 80\% of the known {\sl stellar} population. 
This is to date the largest optical survey of the TMC sensitive well into 
the substellar regime. The 
outline of the survey, overlaid on the CO gas density map, is presented 
in Fig.~\ref{ff1} together with the known TMC population. 

\begin{table*}[t]
\centering
\caption{Overview of the CFHT optical survey of the Taurus cloud}
\begin{tabular}{lccccllc}\hline
Instrument & FOV & Pixel & Date & Area            &      Band & Completeness & Mass limit \\
  & (deg$^2$) & ($^{\prime\prime}$/pixel) & &  (deg$^2$)  &   & limit  & (A$_V$=5, 5Myr)    \\\hline\hline
CFHT12k & 0.33 & 0.21 & 1999-2001 & 3.6 & R, I, z' & 23, 22, 21 & 15$M_{\rm J}$ \\
CFHT12k & 0.33 & 0.21 & 2002 & 8.8 & I, z' & 22, 21 & 15$M_{\rm J}$ \\
Megacam & 1 & 0.19 & 2003-2004 & 34 & i', z' & 24, 23 & 10$M_{\rm J}$ \\\hline
\end{tabular}
\label{tab:phot-survey}   
\end{table*}

The technical characteristics of the CFH12k and MEGACAM cameras are
presented in {\it Cuillandre et al.} (2000) and {\it Boulade et al.} (2003), 
respectively. The first half of the CFH12k survey (centered on the
densest part of TMC, from 1999 to 2001), has been
obtained as part of CFHT director's discretionary time, while the
remaining larger set of data has been obtained, in service mode, as
part of a larger key program devoted to the study of young clusters.
Currently, 28 of the total 34~sq. deg have been reduced and
analyzed. We refer below to this as the primary CFHT survey. 
  Data reduction, performed at CFHT using elements of the
Elixir system ({\it Magnier and Cuillandre,} 2004), included bias and dark
subtraction, flat-fielding, fringing correction, bad pixel removal
and individual frame combination. Point source detection was performed
on the combined $I+z^{\prime}$ images. For the CFH12k data, PSF
fitting photometry was extracted with the PSFex routine from the
SExtractor program ({\it Bertin and Arnouts,} 1996), while aperture photometry was
obtained for the MEGACAM data with the same program. Photometric
catalogs were combined, using the transformation between CFH12k
($I$,$Z^{\prime}$) and MEGACAM ($i^{\prime}$,$z^{\prime}$) photometric
systems, computed with overlapping fields.  The survey yielded more
than $10^6$ sources detected down to $I=24$ and $z^\prime = 23$. From
the turn-over at the faint end of the magnitude distribution, we
estimate the completeness limits of our optical photometric survey to
be $I=21.8$ and $z^\prime=20.9$, which corresponds to a mass completeness limit of 
15~$M_{\rm Jup}$ for $A_V < 5$ and age $<$5~Myr, according
to the pre-main sequence DUSTY models of {\em Chabrier et al.} (2000). On the bright
side, the saturation limits are $i^{\prime} = 12.5$ and $z^{\prime} =12$.
Follow-up studies at NIR wavelengths (1--2~\micron) are planned. As part 
of the UKIRT Infrared Deep Sky Survey most of the MEGACAM fields will be 
observed down to K = 18.

\bigskip
\centerline{\textbf{ 4. OVERVIEW OF THE POPULATION}}
\bigskip

\noindent
\textbf{ 4.1 Fundamental Parameters}\label{fundparam}
\bigskip

We have compiled a comprehensive catalog of all sources thought to be members
of the  TMC, collecting
photometry, effective surface temperature, bolometric luminosities of the stars, $L_{*}$,
derived mostly from near-IR photometry (see, e.g., {\it Brice\~no et al.,} 2002),
extinctions $A_V$ and $A_J$, masses, radii, $H\alpha$ equivalent widths, 
rotation periods and $v\sin i$ values, mass accretion  and  outflow rates, and
some further parameters from the published literature. 

There is considerable spread in some of the photometry and $A_V$ (or $A_J$) estimates for a 
subsample of stars, resulting in notable differences in derived $L_*$ and  masses.
We have coherently re-derived ages and masses from the original $L_*$ and $T_{\rm eff}$
using  {\it Siess et al.} (2000) evolutionary tracks. For the final list of parameters, we have typically 
adopted the values in the recent compilation by {\it Brice\~no et al.} (2002) or, if not available,
the catalogs of {\it Kenyon and Hartmann} (1995) and {\it Brice\~no et al.} (1998). Binary component 
information is mostly from {\it White and Ghez} (2001)
and {\it Hartigan and Kenyon} (2003). Further parameters were complemented from the studies by 
{\it Luhman et al.} (2003), {\it Luhman} (2004), {\it White and Hillenbrand} (2004), 
and {\it Andrews and Williams} (2005).

\bigskip
\noindent{\textbf{ 4.2 Known Protostellar and Stellar Population}}
\bigskip

The bright protostellar and stellar population of
TMC has long been studied in the infrared.
{\it Strom et al.} (1989) concluded that about half of
the young stellar population of TMC above
1~$L_{\odot}$ were surrounded by optically thick
disks as demonstrated by their IRAS detections.
This work was extended and complemented by
the work of {\it Kenyon and Hartmann} (1995) who added
fainter association members and
ground-based photometry out to 5~\micron\ to the
SEDs. ISOCAM observed the L1551 field in the southern
TMC, detecting an additional 15 YSO candidates
({\it Galfalk et al.,} 2004). {\it Spitzer} IRAC photometry
of 82 known Taurus association members is reported
in {\it Hartmann et al.} (2005). This study finds that
the CTTS are cleanly separated from the WTTS in
the [3.6]-[4.5] vs. [5.8]-[8.0] color-color
diagram. The WTTS are tightly clustered around
0 in both colors, and the CTTS form a locus
around [3.6]-[4.5] $\approx$ 0.5 and [5.8]-[8.0]
$\approx$ 0.8.  A similar conclusion is reached
by {\it Padgett et al.} (2006, submitted) who obtained pointed
photometry of 83 WTTS and 7 CTTS in Taurus, Lupus, Chamaeleon, and 
Ophiuchus at distances of about 140 - 180~pc, with ages most
likely around 0.5--3~Myr. They find that
only 6\% of WTTS show excess at 24~\micron, with
a smaller percentage showing IRAC excesses.
Unfortunately, it is currently not possible in
every case to distinguish a true WTTS (pre-main
sequence star without strong H$\alpha$ emission)
from X-ray bright zero-age main-sequence stars
projected onto the cloud. Thus, current samples
of WTTS and possibly ``weak'' BDs may be contaminated with
older objects, skewing the disk frequency for these
sources.

Class I ``protostars" are perhaps more easily
studied in TMC than elsewhere due to the lack
of confusion in the large long wavelength IRAS beams.
One troubling aspect of the placement of Class Is
in the standard picture of star formation ({\it Adams 
et al.} 1987) is that the TMC Class Is
typically show luminosities no higher than, and
in many cases lower than, the Class II T Tauri
stars ({\it Kenyon and Hartmann,} 1995). 
This issue has led to controversy regarding whether Class I
sources are at an earlier evolutionary state than Class II
T Tauri stars ({\em Eisner et al.,} 2005; {\em White and Hillenbrand,} 2004).
It is hoped that {\it Spitzer}
can boost the number of known Class I sources
and elucidate their spectral properties, helping
to determine the true nature of these objects.

\bigskip
\noindent{\textbf{ 4.3 X-Ray Sources}}
\bigskip

The detection statistics of our X-ray survey is summarized in Table~\ref{tbl-1}
(we have added one brown-dwarf detection from a complementary {\it Chandra} field).
An important point for further statistical studies is that the X-ray sample
of detected CTTS and WTTS is nearly complete for the surveyed fields (as
far as the population is known). The few remaining, undetected objects are
either heavily absorbed, have unclear YSO classification, are objects
that have been very poorly studied before, or are very-low mass objects.
Some may not be genuine TMC  members. In contrast, previous X-ray surveys did not detect the 
intrinsically fainter TTS population, potentially introducing bias into statistical 
correlations and population studies. It is little surprising that some of the protostars
remained undetected given their strong photoelectric absorption.  The detection rate of BDs
(53\%) is also very favorable; the remaining objects of this class are
likely to be intrinsically fainter than our detection limit rather than being
excessively absorbed by gas ($A_V$ of those objects typically being no more than 
few magnitudes). 

\begin{table}[h]
\caption{X-ray detection statistics \label{tbl-1} }
\begin{tabular}{lrrr}
\hline
\hline
Object & Members &  Detections & Detection \\  
type   & surveyed    &             & fraction  \\
\hline
Protostars	  & 20  	&    9       & 45\%	\\
CTTS		  & 62  	&   54       & 87\%	\\
WTTS		  & 49         &    48       & 98\%	\\
BDs		  & 19         &    10       & 53\%	\\
\hline
\end{tabular}
\end{table}

X-rays can efficiently be used to find new candidate TMC association members
if X-ray information (luminosity, temporal and spectral characteristics) is combined
with information from the optical/near infrared (placement on the HR diagram, $L_*$, age)
and from the mid-infrared (presence of disks or envelopes). {\it Scelsi et al.} (2006, in preparation)
have thus identified several dozens of potential candidates of the TMC population.
Follow-up studies will be needed to confirm these candidates.

\bigskip
\noindent{\textbf{ 4.4 Bright 24~$\micron$ Sources}}
\bigskip

The scientific goal of the TMC {\it Spitzer} survey is to obtain a 
complete census of the stellar content of these clouds down to the hydrogen
burning limit. Our {\it Spitzer} maps have sufficient sensitivity to
detect 1~M$_{\rm Jup}$ young BDs and optically thin disks
around solar-type stars at the distance of TMC.  However,
a complication of our survey is that the small size of {\it Spitzer}
limits its spatial resolution, making the task of distinguishing
faint stars from galaxies difficult, especially in the presence
of optical extinction. Unfortunately, the IR spectral
energy distributions of extincted stars with infrared
excesses strongly resemble the SEDs of IR-bright galaxies ({\it Evans et al.,} 2003).
Experience with
the galactic First Look Survey ({\it Padgett et al.,} 2004) and the c2d
Legacy program (cf. {\it Young et al.,} 2005) have shown that strong
24~\micron\ emission is an excellent signpost of young stellar
objects. The extragalactic Spitzer surveys performed by the GTOs
and the Extragalactic First Look Survey have established that extragalactic
sources dominate the sky at a flux level of 1 mJy at 24~\micron,
but are fewer than 1 per square degree at 10 mJy ({\it Papovich
et al.,} 2004). Thus, in a region of known star formation,
strong 24~\micron\ sources are more likely to be galactic
than extragalactic sources. Although our analysis of
the TMC {\it Spitzer} maps is incomplete, we have assembled
SEDs for the bright ($\geq$ 10 mJy) 24~\micron\ sources over more than
15 square degrees. 
About 100 sources were found in this
preliminary list, of which 56 have no SIMBAD identifier.
By analogy with SEDs of the known young stellar objects
of the cloud which were also recovered by this technique,
we believe that some of the previously unknown 24~\micron\ sources may represent
the brightest stars with disks among the YSOs which were too faint for IRAS 
to detect. SEDs for four of the new bright 24~\micron\ sources
are presented in Fig.~\ref{ff3}.

\begin{figure}[h!]
\epsscale{1.0}
\centerline{\resizebox{0.9\hsize}{!}{\includegraphics{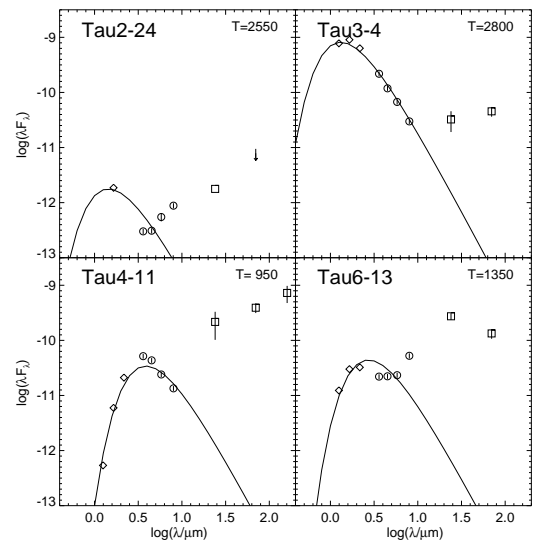}}}
\caption{\baselineskip 13pt \small 2MASS NIR + {\it Spitzer} IRAC and MIPS
spectral energy distributions of four bright 24 \micron\ sources
discovered in the course of the survey. Temperatures indicated in the plots are
effective temperatures of the plotted photosphere.
\label{ff3}}
\end{figure}

\bigskip
\centerline{\textbf{ 5. THE SUBSTELLAR SAMPLE}}
\bigskip

The TMC is a particularly interesting target for
searches of young substellar objects.  
It has a large
extension, so it can be studied for ejection effects; there are no
bright stars there to irradiate and disturb the stellar surroundings;
the census of stellar members is relatively complete down to M2V spectral
types ({\it Kenyon \& Hartmann,} 1995), and its 
spatial distribution is known ({\it G\'omez et al.,}  1993). The average low
extinction ($A_V = 1$) associated with this cloud as well as its young age
combine to provide a high sensitivity to very low-mass objects in the
optical domain.  However, its large spatial extent ($\simeq$ 100
square degrees) requires mapping an extensive area. Significant
breakthrough in this domain has been made possible with the recent
availability of large scale optical cameras.

Searches for substellar objects, by {\it Brice\~no et al.} (1998)
{\it Luhman} (2000), {\it Brice\~no et al.} (2002), {\it Luhman et al.} (2003), {\it Luhman} (2004) 
and references therein, have revealed a factor of 1.4 to 1.8 deficit of
BDs with respect to stars in TMC compared to the
Trapezium cluster. This result has been interpreted as an indication
that sub-stellar object formation depends on the environment.
However, all these previous studies were concentrated on the immediate
vicinity of the high stellar density regions. If BDs are stellar
embryos ejected from their birth sites early in their evolution as
proposed by {\it Reipurth and Clarke} (2001), a significant fraction of the
substellar content of the central parts of the cloud could have
scattered away and may have been missed. This being the main scientific
driver of the CFHT survey, we describe recent CFHT results from the search
for substellar objects in TMC based on {\it Mart\'{\i}n et al.} (2001) 
and  {\it Guieu et al.} (2006) below, together with aspects from {\it Spitzer} 
and {\it XMM-Newton}.

\bigskip 
\noindent 
\textbf{ 5.1 New TMC Very Low-Mass Members} 
\bigskip 
 
\begin{figure}[t] 
\centerline{\resizebox{0.80\hsize}{!}{\includegraphics{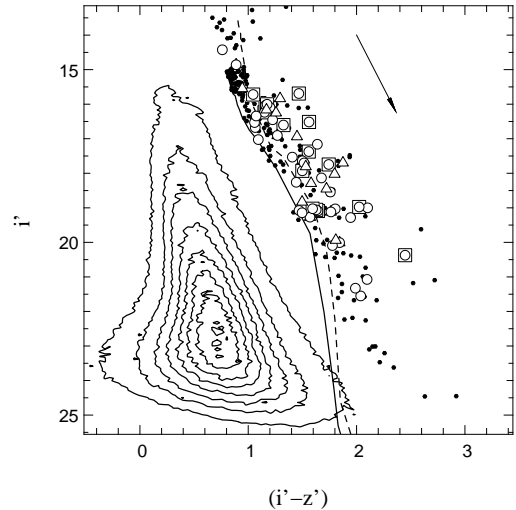}}}
\caption{Observed $i^{\prime}/(i^{\prime}-z^{\prime})$ color-magnitude
diagram used to select low-mass TMC candidates. Small black dots
are candidate TMC members. The photometric mid to late-M candidates
observed spectroscopically by {\it Guieu et al.} (2006) are displayed
by black open circles. Triangles are previously known TMC
members. Squares   identify the spectroscopically confirmed 21 new
TMC members from {\it Guieu et al.} (2006) and {\it Mart\'{\i}n et al.} (2001). The 
two steep solid curves show the locations of the 1 Myr and 10 Myr isochrones from 
the DUSTY model of {\it Chabrier et al.} (2000) at the TMC distance.  The arrow 
indicates a reddening vector of $A_V$ = 4 magnitudes. Figure adapted 
from {\it Guieu et al.} (2006).}      
\label{ff4}   
\end{figure}

Substellar photometric candidates are identified from their location
in combined optical/NIR color-magnitude and color-color diagrams (see
Fig.~\ref{ff4}). The full details of the selection process are
given in {\it Guieu et al.} (2006). Complementary near-infrared photometry,
taken from the 2MASS catalog, is critical to reduce the strong
expected galactic contamination, primarily from background giants.
Residual contamination is still expected to be at the 50 \% level. In
order to properly assess TMC membership, spectroscopic follow-up of
the photometrically selected candidates is therefore mandatory.  The
criteria used to assess TMC membership are detailed in
{\it Guieu et al.} (2006). They rely on estimates of the surface gravity, obtained
both from spectral fitting and measurements of the Na {\sc i} equivalent
widths.  The level of H${\alpha}$ emission is used as an additional
indicator of youth.  At the median age of the TMC population of 3
Myr, the pre-main sequence models of {\it Chabrier et al.} (2000) predict the
stellar/substellar boundary to lie at a spectral type between M6 and
M6.5V.

The photometric selection procedure yielded, over the primary 28
sq. degrees CFHT survey, 37 TMC mid- to late-M spectral type
new candidate members with $i^{\prime} < 20$ (magnitude limit set
to allow a proper spectral type determination).  TMC membership has
been confirmed spectroscopically for 21 of these sources
({\it Mart\'{\i}n et al.} 2001, {\it Guieu et al.} 2006), 16 of which have spectral
types later than M6.5V, i.e. are likely substellar. These new findings
bring to 33 the current published census of TMC BDs, thus
allowing for a preliminary statistical study of their properties.

\bigskip 
\noindent 
\textbf{ 5.2 A Peculiar Substellar IMF in TMC?} 
\bigskip 

When all published optical surveys are combined,
the census for very low-mass TMC objects is now complete down to
30~$M_{\rm Jup}$ and $A_V \leq 4$ over an effective surface area of 35
sq. deg ($\approx 30$\% of the total cloud surface). This
mass completeness, derived from the DUSTY model of
{\it Chabrier et al.} (2000), is set both by the 2MASS completeness limits
(J=15.25, H=14.4, K=13.75) and the typical sensitivity limit of
optical spectroscopic observations (i$^{\prime}\leq$20).

{\it Brice\~no et al.} (2002) have introduced the substellar-to-stellar ratio
$$R_{ss} = {{N(0.03 <M/M_\odot<0.08)}\over {N(0.08<M/M_\odot<10)}}$$
as a measure of the relative abundance of BDs. In their
pioneering study, targeted towards the high-density aggregates,
{\it Brice\~no et al.} (2002)  determined a value of $R_{ss}$ ($0.13 \pm 0.05$)  
lower by a factor 2 than the one found in the Trapezium cluster.  This
study thus suggested that the relative abundance of BDs in
star-forming regions may depend on initial conditions, such as the
stellar density. However, in a recent, more extensive study covering
12 square degrees, {\it Luhman} (2004) found that the BD
deficit in TMC (with respect to Orion) could be less pronounced
than previously thought.  Combining the recent new members from the
CFHT~survey with previously published results, {\it Guieu et al.} (2006)
derive an updated substellar-to-stellar ratio in TMC of $R_{ss} =
0.23 \pm 0.05$. This value is now in close agreement with the
Trapezium value of $R_{ss} (\mbox{Trapezium}) = 0.26\pm 0.04$
estimated by {\it Brice\~no et al.} (2002), using the same evolutionary models
and treating binary systems in the same manner. It also appears
consistent with the more recent values derived for IC~348 by
{\it Slesnick et al.} (2004), the Pleiades by
{\it Moraux et al.} (2003) and computed from the galactic disk system IMF of
{\it Chabrier} (2003) (see {\it Monin et al.,} 2005, for a compilation of these 
values). These new findings seem to suggest a universal 20--25\% value for the
relative abundance of BDs in young clusters. The fact that
the estimate of the substellar-to-stellar ratio in TMC has kept
increasing as larger areas were surveyed suggests that this ratio may
depend on the local stellar density. Indeed,  {\it Guieu et al.} (2006) find
evidence for a deficit of the abundance of BDs of a factor $\simeq$ 2
in the central regions of the TMC aggregates (on scales of $\simeq$
0.5 pc) with respect to the more distributed population. 
As discussed in {\em Guieu et al.} (2006), this result may be an indication
for spatial segregation resulting from ejection of the lowest-mass
members and seem to favor dynamical evolution of small N body 
systems as the formation process of substellar objects (see {\em Guieu et al.} 
2006 for a full discussion).

\bigskip 
\noindent 
\textbf{ 5.3 X-Ray Properties of TMC BDs} 
\bigskip 

The area surveyed by the 28 pointings of {\it XMM-Newton}, 
combined with one {\it Chandra} archival observation, allow us to
study the X ray emission of 19/33 TMC BDs ({\it Grosso et al.,}
2006a). Among these, ten BDs are detected, yielding a detection rate of 
$\approx 53\,$\%. One BD displayed an impulsive flare, demonstrating 
variability in X-rays over periods of a few hours.
The detection rate of TMC BDs thus appears similar to the
one in Orion where it reaches $\approx 50$\% for $A_V<5$ ({\it Preibisch
et al.,} 2005b). As in Orion, there is a tendency to detect earlier
(hotter) BD, with spectral types earlier than M7-M8. 

\begin{figure}[h!]
\centerline{\rotatebox{90}{\resizebox{0.6\hsize}{!}{\includegraphics{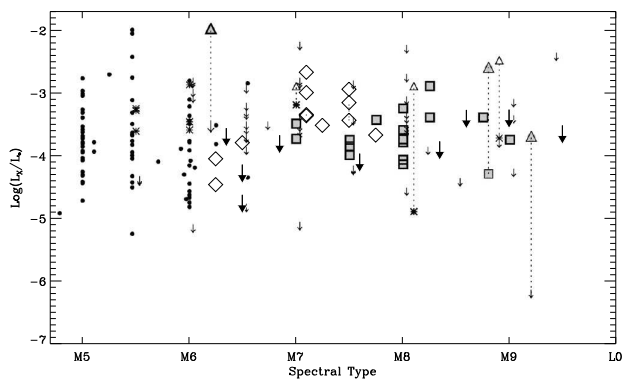}}}}
\caption{\small $L_X/L_{\star}$ of low-mass stars and BDs as a function
of spectral class, including samples of late-type main-sequence
field stars (asterisks, {\it Fleming et al.,} 1993), the Orion Nebula Cluster BD sample
(filled squares, triangles, and small arrows)  and T Tauri stars later than
M5 (filled dots) from {\it Preibisch et al.} (2005b),
and our sample of TMC detections (diamonds) and upper limits (arrows). (Figure courtesy of Nicolas Grosso.)
 }\label{ff5} 
\end{figure}

There is appreciable scatter in the X-ray luminosities. The most luminous examples show 
$L_X$ of order $\approx 10^{29}$~erg~s$^{-1}$.
No trend is seen for $L_X/L_{*}$ with spectral type, i.e., the efficiency of magnetic 
field production and coronal heating appears to be constant in low-mass stellar and substellar
objects (Fig.~\ref{ff5}). 

\bigskip 
\noindent 
\textbf{5.4 Disk and Accretion Properties of TMC BDs} 
\bigskip 
 
There is now ample evidence that TMC BDs experience accretion
processes similar to the more massive TTS.
Near-infrared L band excesses have been detected in TMC substellar
sources, indicating a disk frequency of $\approx 50$\% ({\it Liu et
al.,} 2003; {\it Jayawardhana et al.,} 2003).  Broad asymmetric $H\alpha$
emission profiles characteristic of accretion are reported in a few
TMC BDs ({\it Jayawardhana et al.,} 2002, 2003; {\it White and Basri,}
2003; {\it  Muzerolle et al.,}  2005, and references therein; {\em Mohanty et
al.,} 2005). Extending the study of {\it Barrado y Navascues and Mart\'{\i}n} (2003), 
{\it Guieu et al.} (2006) find that the fraction of BDs in TMC with levels of
H$\alpha$ emission in excess of chromospheric activity to be 42\%, similar
to the low-mass TTS. 

Of the twelve BD candidates optically selected from the
CFHT survey, half have strong excesses in the
infrared as measured from {\it Spitzer} data ({\it Guieu et al.} 2005).
Most of these diverge from the predicted photospheric fluxes
at 5.8 \micron, and all six are strongly detected at
24 \micron. The other substellar candidates have IRAC
fluxes indistinguishable from a late-M photosphere,
and only one is detected at 24 \micron. These results
are similar to those found for TMC BD in the literature
by {\it Hartmann et al.} (2005). Both studies find that the
BD with excess (``Classical" BD or CBD) have disk properties
indistinguishable from classical T Tauri stars. Similarly,
the ``weak" BDs  have purely photospheric
colors similar to the WTTS. Further analysis and modeling
is required to determine whether the ``classical'' BDs have
unusually flat disks as suggested by {\it Apai et al.} (2002).

\begin{figure}[h!]
\centerline{\resizebox{0.85\hsize}{!}{\includegraphics{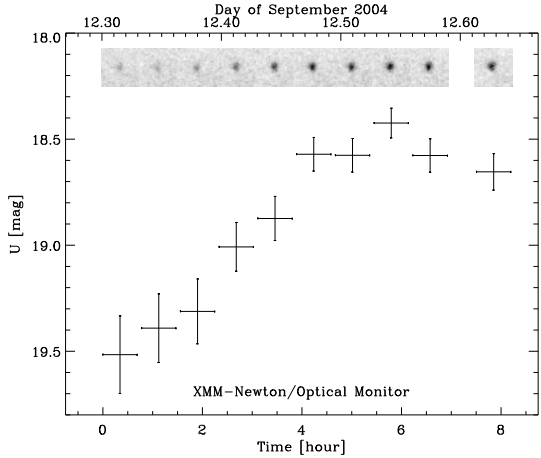}}}
\caption{\small  U-band light curve of 2MASS~J04552333+3027366  during an {\it XMM-Newton} 
observation. The slow increase may be ascribed to an accretion event. The insets show U-band 
images from which the  fluxes have been extracted. Background limiting magnitudes are 
indicated by thick horizontal lines ({\it Grosso et al.}, 2006b).
 }\label{ff6}%
\end{figure}

There is support for BD variability in the U-band observations obtained simultaneously
by the Optical Monitor (OM) onboard {\it XMM-Newton}. 
The OM observed  13 of the 19 X-ray surveyed BDs in the U-band. Only one BD 
was detected, 2MASS~J04552333+3027366, for which the U band flux 
increased  by a factor of about 2--3 in $\approx$6~hours (Fig.~\ref{ff6};
{\it Grosso et al.,} 2006b).  The origin of
this behavior can be explained by several different mechanisms: either
rotational modulation of a hot or dark spot, or a coronal magnetic flare,
or variable accretion. This BD was not detected in X-rays at any time.
It is known to accrete at a rate of $10^{-10}~M_{\odot}$~yr$^{-1}$ 
({\it Muzerolle et al.,} 2005).
Assuming that the relation $\log L_{\rm acc} \propto \log L_{\rm U}$ that
applies to TTS ({\it Gullbring et al.,} 1998) is also valid for BDs,
the accretion rate must have increased by a factor of about 2--3 during
the observing time to explain the increase observed in the U-band.

 All these results argue for a
continuous $M/\dot{M}$ relation through the stellar/substellar
boundary, as illustrated e.g. in Fig.~5 of {\it Muzerolle et al.} (2005).

\bigskip 
\noindent 
\textbf{5.5 Implication for Substellar Formation Model} 
\bigskip 

 The fact that the abundance of BDs (down to 30 M$_{\rm
 Jup}$) relative to stars is found to be the same ($\simeq 25$\%) in
 the diffuse TMC {\em and} in the high-density Orion Nebula Cluster 
 seems to suggest that there is no strong dependency of the substellar
 IMF on initial molecular cloud conditions, in particular gas density
 and level of turbulence.  This fact and the increase of the BD
 abundance with decreasing stellar density found by {\it Guieu et al.} (2006)
 could be best explained if a fraction of the distributed population
 in TMC is formed of low-mass stars and substellar objects ejected
 from the aggregates through rapid dynamical decay in unstable small
 N-body systems ({\sl the ejected-embryo} model). Indeed, such a result
 is predicted by the dynamical evolution studies of {\it Kroupa and Bouvier}
 (2003) and the recent sub-sonic turbulent fragmentation simulations
 of {\it Goodwin et al.} (2004). For a detailed discussion, see
 {\it Guieu et al.} (2006) and {\it Monin et al.} (2006).

 It has often been argued that the presence of accretion / outflow
 activity in BDs would be incompatible with an ejection formation
 scenario.  However, even truncated disks in ejected objects can
 survive for a few Myr, a period consistent with the TMC age. The
 viscous timescale of a disk around a central mass $M$ varies as
 $M^{-1/2}$, so for a disk truncated at $R_{\rm out}=10\,$AU,
 $\tau_{\rm visc}\approx 2\,$Myr around a 50\,$M_{\rm Jup}$ BD (for
 $\alpha\approx 10^{-3}$ at 10\,AU). Furthermore, an accretion
 rate of $\dot{M}=10^{-11}\,M_\odot {\rm yr}^{-1}$ and a disk mass of
 $10^{-4}\,M_\odot$ results in a similar lifetime of a few Myr. So,
 there is no contradiction between BD ejection and the
 presence of (possibly small) accretion disks at a few Myr age.

\bigskip
\centerline{\textbf{ 6. X-RAYS AND MAGNETIC ACTIVITY}}
\bigskip

\noindent
\textbf{ 6.1 X-Ray Luminosity}\label{sect-6.2}
\bigskip

\begin{figure}[b!]
\resizebox{0.95\hsize}{!}{\includegraphics{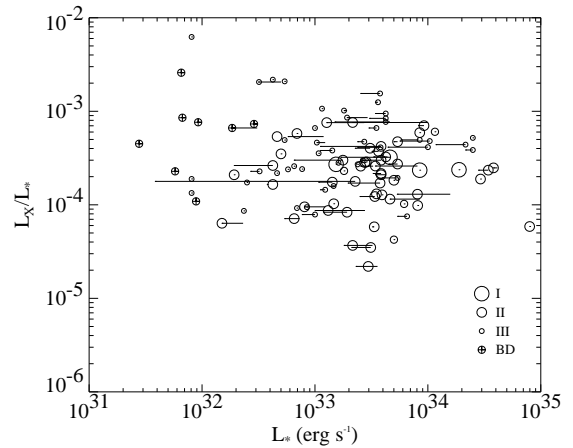}}
\vskip -0.2truecm
\caption{\small Plot of $L_X/L_*$ as a function of $L_{*}$ for all X-ray detected (and spectrally
modeled) stars and BDs (but excluding Herbig stars). 
Symbol size, from largest to smallest: protostars (IR Class I) -  CTTS (or IR Class II) - 
WTTS (or IR Class III). 
Circles with crosses: BDs. The error bars indicate ranges 
of $L_{*}$ given in the literature. 
 }\label{ff7}%
\end{figure}

\begin{figure}[h!]
\resizebox{0.95\hsize}{!}{\includegraphics{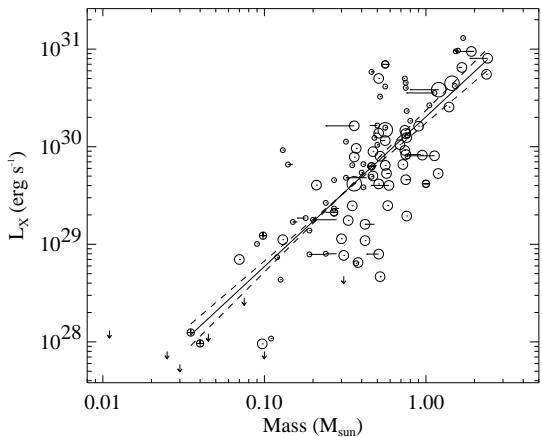}}
\vskip -0.4truecm
\caption{\small X-ray luminosity $L_X$ versus stellar mass (excluding Herbig stars). 
A clear correlation is visible. Key to the symbols is as in Fig.~\ref{ff7} 
({\em after Telleschi et al.,} 2006, in preparation).  
 }\label{ff8}%
\end{figure}

Fig.~\ref{ff7} shows the distribution of the ratio between X-ray luminosity $L_X$ 
and (stellar, photospheric) bolometric luminosity $L_{*}$ as a function of 
 $L_{*}$ (the latter derived from optical or near-IR data, see Section~4.1 for
references) for all spectrally modeled TTS and protostars,  and also including the 
{\it detected} BDs (we exclude the peculiar sources discussed in Section~6 below; 
some objects were observed twice with different $L_X$ - we use the logarithmic averages of $L_X$ in these cases). 
We do not give errors for $L_X$ because most objects are variable on short and long  time 
scales (hours to days), typically within a factor of two outside obvious, outstanding flares.
Most stars cluster between $L_X/L_{*} = 10^{-4} - 10^{-3}$ as is often found in star-forming regions 
({\it G\"udel} 2004 and references therein). The value $L_X/L_{*} = 10^{-3}$ corresponds
to the saturation value for rapidly rotating main-sequence stars (see below). We  note a
trend for somewhat lower levels of $L_X/L_{*}$ for higher $L_{*}$ (typically, more 
massive stars). What controls the X-ray luminosity level? Given the trend toward saturation
in Fig.~\ref{ff7}, one key parameter is obviously $L_{*}$. Although for pre-main sequence
stars there is no strict correlation between $L_{*}$ and stellar mass, it is interesting
that we find a rather well-developed correlation between $L_X$ and mass $M$ (Fig.~\ref{ff8},
masses derived from $T_{\rm eff}$ and $L_*$ based on {\it Siess et al.} 2000 isochrones)
that has been similarly noted in Orion ({\it Preibisch et al.,} 2005a).
Part of this correlation might be explained by higher-mass stars being larger, i.e., 
providing more surface area for coronal active regions. The correlation between surface
area and $L_X$ is, however, considerably weaker  than the trend shown in Fig.~\ref{ff8}.

\begin{figure}[h!]
\resizebox{0.85\hsize}{!}{\includegraphics{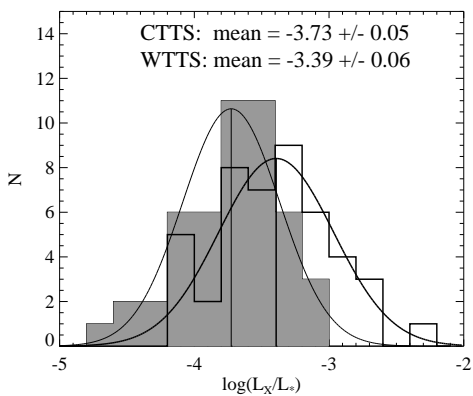}}
\vskip -0.4truecm\caption{\small Comparison of the $L_X/L_{*}$ distributions for CTTS
(shaded histogram) and WTTS (solid), together with log-normal fits. The CTTS sample is on 
average less luminous (normalized to $L_{*}$) than the WTTS sample. The errors in 
the plot indicate the error of the means of the distributions ({\em Telleschi et al.,} 2006, in preparation). 
 }\label{ff9}%
\end{figure}

We plot in Fig.~\ref{ff9} the $L_X/L_{*}$  distribution separately
for CTTS and WTTS ({\it Telleschi et al.,} 2006, in preparation; the average  $L_X$ is used for objects 
observed twice).  Because our samples are nearly complete, there is 
little bias by detection limits. The distributions are close to log-normal, and
corresponding Gaussian fits reveal that WTTS
are on average more X-ray luminous (mean of distribution: 
$\log L_X/L_{*} = -3.39\pm 0.06$)
than CTTS (mean: $\log L_X/L_{*} = -3.73\pm 0.05$), although the widths of 
the distributions are similar. This finding parallels earlier reports on
less complete samples ({\it Stelzer and Neuh\"auser,} 2001), ruling out detection bias 
as a cause for this difference. A similar segregation into two X-ray populations
has not been identified in most other SFRs (e.g., {\it Preibisch and Zinnecker,} 2001 - but see
recent results on the Orion Nebula Cluster in {\it Preibisch et al.,} 2005a). The cause 
of the difference seen in TMC may be evolutionary (stellar size, convection zone depth), 
or related  to the presence of accretion disks or the accretion process itself. We will 
return to this point in Section~6.3 below.

\bigskip
\noindent
\textbf{ 6.2 Rotation and Activity} 
\bigskip

Rotation plays a pivotal role for the production of magnetic fields in main-sequence
stars, and thus for the production of ionizing (ultraviolet and X-ray) radiation. The rotation 
period $P$ is controlled by the angular momentum of the young star inherited from the contracting 
molecular cloud, by the further contraction of the star, but possibly also by magnetic fields
that connect the star to the inner border of the circumstellar disk  and thus apply
torques to the star. Strictly speaking, in the standard (solar) $\alpha-\omega$ dynamo theory,
it is {\it differential} rotation that, together with convection, produces magnetic flux near
the boundary between the  convection zone and the  radiative core. Because
the younger T Tau stars are fully convective, a dynamo of this kind is not expected, but
alternative dynamo theories based entirely on convective motion have been proposed
(e.g., {\em Durney et al.,} 1993). It is
therefore of prime interest to understand the behavior of a well-defined sample of
T Tau stars. 

In cool  main-sequence stars, a rotation-activity relation is found for $P$
exceeding a few days (the limit being somewhat dependent on the stellar mass or spectral type),
approximately following $L_X \propto P^{-2.6}$ ({\it G\"udel et al.,} 1997, {\it Flaccomio et al.,} 2003a). Given
the role of the convective motion, a better independent variable may be the Rossby number $R = P/\tau$ 
where $\tau$ is the convective turnover time ({\em Noyes et al.,} 1984). If the rotation period is smaller than a few days, 
the X-ray luminosity saturates at a value of $L_X/L_{*} \approx 10^{-3}$ and stays at this level
down to very short periods.

Corresponding studies of TTS have produced conflicting results. Although 
a  relation has been indicated in TMC ({\it Stelzer and Neuh\"auser,} 2001),
samples in other star-forming regions show
stars at saturated X-ray luminosities all the way to periods of $\approx 20$~days (e.g.,
{\it Preibisch et al.} 2005a). There is speculation that these stars are still within the saturation 
regime because their Rossby number remains small enough for the entire range of $P$, given 
the long convective turnover times in fully convective stars.

Our nearly complete sample of TTS (for the surveyed area) permits an unbiased investigation of
this question with the restriction that we  know $P$ for
only 25 TTS (13 CTTS and 12 WTTS) in our sample. Another 23 stars 
(15 CTTS and 8 WTTS) have measured projected rotational velocities $v\sin i$ which imply
upper limits to $P$ once the stellar radius is known. In a statistical sample with random 
orientation of the rotation axes, the average of $\sin i$ is $\pi/4$ which we used for 
estimates of $P$ if only $v\sin i$ was known. The stellar radii were calculated from
$T_{\rm eff}$ and the (stellar) $L_{*}$.

\begin{figure}[h!]
\hskip -0.5truecm\hbox{
\resizebox{0.9\hsize}{!}{\includegraphics{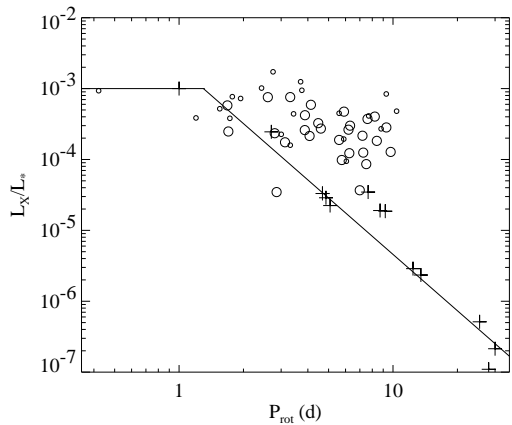}} 
}
\vskip -0.5truecm
\caption{\small 
The ratio $L_X/L_{*}$ as a function of rotation period for the TMC sample ({\em Briggs et al.,} 2006, in preparation). 
Symbols are as in Fig.~\ref{ff7}.
The crosses and the schematic power-law fit resp. horizontal saturation
law apply to a sample of solar analogs on the main sequence ({\it G\"udel et al.,} 1997).
 }\label{ff10}%
\end{figure}

The resulting trend is shown in Fig.~\ref{ff10} (the average $L_X$ is used for stars with 
multiple observations; {\it Briggs et al.} 2006, in preparation).
 First, it is evident that the sample of CTTS with measured $P$ rotates, on average, less 
 rapidly than WTTS  (characteristically, $P \approx$ 8~d and 4~d, respectively).  
Fig.~\ref{ff10} shows that the rotation-activity  
behavior is clearly different from that of main-sequence solar-mass stars in that 
$L_X/L_{*}$ remains at a saturation level up to longer periods. This is not
entirely surprising given that the same is true for less massive main-sequence K and M-type
stars that are more representative of the TTS sample ({\it Pizzolato et al.,} 2003).
The trend is even clearer when plotting the average X-ray surface flux, in particular for 
periods exceeding $\approx 5$~d ({\em Briggs et al.,} 2006, in preparation), supporting previous ROSAT 
studies ({\it Stelzer and Neuh\"auser,} 2001).

Why this finding is at variance from findings in Orion ({\it Preibisch et al.,} 2005a) is unclear.
One possibility are unknown survey biases ({\it Briggs et al.}, 2006, in preparation). Another reason is the (on average) larger
age of TMC in which a larger fraction of stars may have developed a radiative core ({\em Briggs et al.}, 2006, 
in preparation).

\bigskip
\noindent
\textbf{ 6.3 Accretion and Disks}\label{accdisk}
\bigskip

In the standard dynamo interpretation, the (on average) slower rotation of the CTTS compared to
WTTS and their (on average) slightly lower $L_X$  are well explained
by a decreasing dynamo efficiency with decreasing rotation rate. This is the conventional
explanation for the activity-rotation relation in aging main-sequence stars.
The  relation suggested above could, however,  be mimicked by 
the CTTS sample, rotating less rapidly, being subject to suppressed X-ray production for
another reason than the decreasing efficiency of the rotation-induced dynamo. We already 
found that the average $L_X/L_{*}$ is smaller by a factor of two for
CTTS compared to WTTS (Section~6.1). The most obvious distinction between CTTS and
WTTS is active accretion from the disk to the star for the former class.

\begin{figure}[h!]
\hskip -0.3truecm\hbox{
\resizebox{0.9\hsize}{!}{\includegraphics{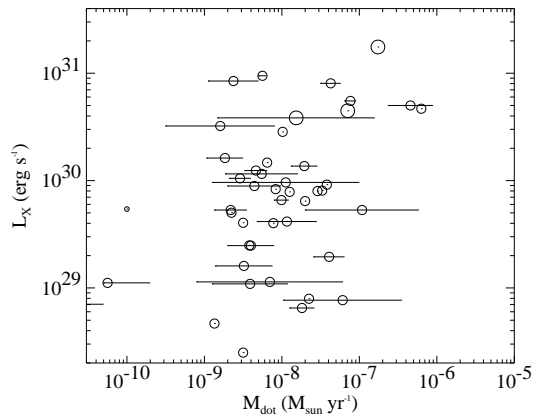}} 
}
\vskip -0.5truecm
\caption{\small 
Scatter plot of $L_X$ 
versus the (range of) mass accretion rates reported in the literature. 
No trend is evident. Symbols are as in Fig.~\ref{ff7}. ({\it Telleschi et al.} 2006, in preparation.) }
	   \label{ff11}%
\end{figure}

There are two arguments against this explanation. First, a rotation-activity relation holds
{\it within} the CTTS sample, and there is no obvious correlation between $P$ and the mass accretion rate, $\dot{M}$, for that
sample. And second, when investigating the coronal properties $L_X$, $L_X/L_{*}$ (and also average
coronal temperature $T_{\rm av}$) as a function of the mass accretion rate (as given by {\it White and Ghez,} 2001;
{\it Hartigan and Kenyon,} 2003; {\it White and Hillenbrand,} 2004), we see no trend
over three orders of magnitude in $\dot{M}$ (Fig.~\ref{ff11} for $L_X$; {\it Telleschi et al.,} 2006, in preparation).  
Mass accretion rate does therefore not seem to be a sensitive parameter that determines overall 
X-ray coronal properties. It therefore rather seems that CTTS produce, on average, lower $L_X$ 
because they are typically rotating more slowly, which may be related to disk-locked rotation
enforced by star-disk magnetic fields (e.g., {\it Montmerle et al.,} 2000).

\bigskip
\centerline{\textbf{ 7. JETS AND OUTFLOWS}}
\bigskip

Shock speeds in the high-velocity component of protostellar jets 
may be sufficient to shock-heat plasma to X-ray temperatures. 
The shock temperature is $T\approx 1.5\times 10^5v_{\rm 100}^2$~K where
$v_{100}$ is the shock front speed relative to a target, in units of 100~km~s$^{-1}$
({\it Raga et al.} 2002). Jet speeds in TMC are typically of order
$v = 300-400$~km~s$^{-1}$ ({\it Eisl\"offel and Mundt,} 1998; {\it Anglada,} 1995; {\it Bally et al.,} 2003), 
allowing for shock speeds of similar magnitude. If a flow shocks 
a standing medium at 400~km~s$^{-1}$, then $T \approx 2.4$~MK. X-rays have been 
detected from the L1551 IRS-5 protostellar jet about 0.5--1$^{\prime\prime}$ away
from the protostar, while the central star is entirely 
absorbed by molecular gas ({\it Bally et al.,} 2003).

X-rays cannot be traced down to the acceleration region or the collimation region
of most protostellar jets because of the considerable photoelectric absorption in particular
of the very soft X-ray photons expected from shocks (energy $\la 0.5$~keV).  An interesting 
alternative is provided by the study of strong jets and micro-jets driven by optically
revealed T Tau stars. {\it Hirth et al.} (1997) surveyed TMC CTTS for evidence of outflows and microjets
on the 1$^{\prime\prime}$ scale, identifying low-velocity (tens of km~s$^{-1}$) and high-velocity 
(up to hundreds of km~s$^{-1}$) flow components in several of them. 

\begin{figure}[h!]
\hbox{
\rotatebox{270}{\resizebox{0.65\hsize}{!}{\includegraphics{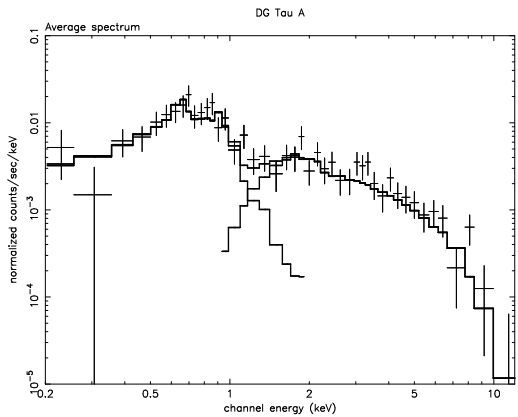}}} 
}
\caption{\small Average spectrum of DG Tau A. 
Also shown is the fit to the spectrum (black histogram) and its two constituents, 
the soft and the hard components. ({\it G\"udel et al.,} 2006b.) }
\label{ff12}%
\end{figure}

X-ray observations of these jet-driving CTTS have  revealed 
new X-ray spectral phenomenology in at least three, and probably four, of these objects in TMC 
(DG Tau A, GV Tau A, DP Tau, and tentatively  CW Tau - see Fig.~\ref{ff12}; {\it G\"udel et al.,} 2005;
{\it G\"udel et al.,}  2006b).  They share X-ray 
spectra that are composed of two different emission components  subject to entirely different 
photoelectric absorption.  The soft component,
subject to very low absorption ($N_{\rm H} \approx 10^{21}$~cm$^{-2}$), 
peaks at 0.7--0.8 keV where Fe~XVII produces strong emission lines,
suggestive of low  temperatures. This is borne out by spectral 
modeling, indicating temperatures of 2--5~MK. Such temperatures are not common to TTS.
A much harder but strongly absorbed component ($N_{\rm H}$ several times 
$10^{22}$~cm$^{-2}$) indicates extremely hot (several tens of MK) plasma.

\begin{figure}[t!]
\hbox{
\resizebox{0.80\hsize}{!}{\includegraphics{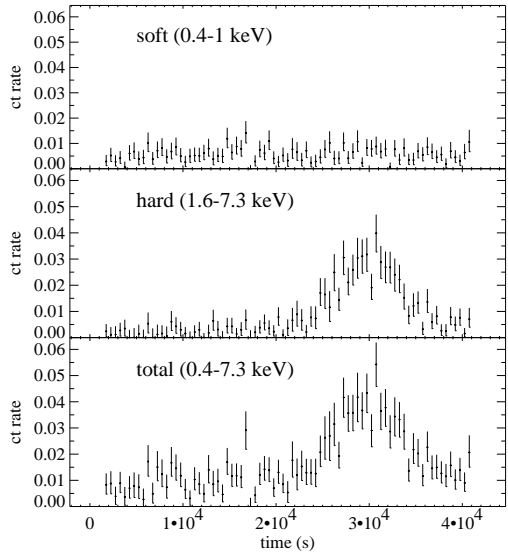}} 
}
\caption{\small X-ray light curves of DG Tau A. The top panel shows the soft 
component, the middle panel the hard component, and the bottom the total X-ray light curve.
({\it G\"udel et al.,} 2006b.) }
\label{ff13}%
\end{figure}

These objects show flares in their X-ray light curves 
(Fig.~\ref{ff13}), but such variability is so
far seen only in the hard component while the soft component is steady. 
Evidently, these ``two-absorber'' spectra require that {\it two physically unrelated 
X-ray sources are present around these objects.}

All these stars are strong accretors ($\dot{M}$ of order $10^{-7}$ to $10^{-6}$~M$_{\odot}$~yr$^{-1}$). 
However, as shown above, the TMC sample reveals  no significant relation between mass 
accretion rate and coronal properties. The distinguishing 
property of these objects is, in contrast, the presence of well-developed, protostar-like 
jets and outflows with appreciable mass-loss rates ($10^{-7}$ to $10^{-6}$~M$_{\odot}$~yr$^{-1}$).

A tentative interpretation is the following
({\it G\"udel et al.,} 2005;   2006b): The flaring
in the hard component occurs on timescales of hours, suggesting ordinary
coronal active regions. The preceding U-band bursts signal the initial 
chromospheric heating before plasma is evaporated into the coronal magnetic loops. The
flaring active regions are therefore likely to be of modest size, well connected to
the surface active regions. The excess absorption is probably
due to cool gas that streams in from the disk along the magnetic field lines, 
enshrouding the magnetosphere with absorbing material. This increases the photoelectric absorption
of X-rays but does not 
increase the optical extinction because the gas streams are very likely to be depleted of dust 
(the latter being evaporated farther away from the star). As for the cool X-ray component, although its 
temperature is also compatible with shock heating of material in accretion columns close to 
the star (e.g.,  {\it Kastner et al.,} 2002), the low photoabsorption makes this interpretation 
problematic and prefers a location outside the magnetosphere. An obvious location of the 
cool, soft  X-ray sources are shocks forming near the base or the collimation region of the jet 
(e.g., {\it Bally et al.} 2003). Jet speeds of several hundred km~s$^{-1}$  support
this model, as do estimated X-ray luminosities (see {\it G\"udel et al.,} 2005, based on the
theory of {\it Raga et al.,} 2002).

If this model is correct, then the consequences are far-reaching:  distributed, large-scale
X-ray sources may efficiently ionize larger parts of the circumstellar environment than
the central star alone, and in particular the disk surface, thus inducing disk accretion 
instabilities ({\it Balbus and Hawley} 1991) and altering the disk chemistry  
({\it Feigelson and Montmerle,} 1999; {\it Glassgold et al.,} 2004).  

\bigskip
\centerline{\textbf{ 8. SUMMARY}}
\bigskip

The Taurus Molecular Cloud provides unequaled insight into the detailed physical
processes of low-mass star formation, environmental issues, astrochemistry aspects, 
and evolutionary scenarios down to the substellar level. New observatories
now available help us tackle outstanding problems with unprecedented sensitivity, 
spectral and spatial resolution. Of particular interest to star-formation studies 
are the new X-ray observatories ({\it XMM-Newton} and {\it Chandra}), the {\it Spitzer} 
Space Telescope in the infrared, and deep, large-scale optical surveys such as the CFHT
survey summarized here.

Combining the X-ray, infrared, and optical population studies, there is considerable potential
for detection of new Taurus members, some of which may be strongly embedded or extincted
by their disks. Joint multi-wavelength studies have been particularly fruitful for
the characterization of brown dwarfs, which have been amply
detected by all three studies and are now supporting a model in which a fraction of
these objects are ejected from denser stellar aggregates.

The surveys  also deepen previous studies of properties of
T Tau stars and protostars (e.g., rotation-activity relations, disk
properties, etc), while at the same time opening the window to new types
of phenomenology such as accretion events on brown dwarfs or X-ray emission
perhaps forming at the base of accelerating jets. Further insight is 
expected to be obtained from high-resolution (optical, IR, and X-ray)
spectroscopy that should probe composition and structure of accretion disks and
heated X-ray sources. 

\bigskip

\textbf{ Acknowledgments.} We thank our referee for constructive and helpful 
comments on our paper. We acknowledge extensive contributions to this 
work by the three TMC teams ({\it XMM-Newton}, {\it Spitzer}, and CFHT).  
M. G\"udel specifically acknowledges help from K. Arzner, M. Audard, K. Briggs, E. Franciosini, 
N. Grosso (particularly also for the preparation of Fig. 1), G. Micela, 
F. Palla, I. Pillitteri, L. Rebull, L. Scelsi, and A.  Telleschi.
D. Padgett thanks J. Bouvier, T. Brooke, N. Evans, P. Harvey, 
J.-L. Monin, K. Stapelfeldt, and S. Strom for contributions.
C. Dougados wishes to thank the whole CFHT TMC survey team, in
particular F. M\'enard, J.~L. Monin, S. Guieu, E. Magnier, E. Mart\'{\i}n and
the CFHT astronomers (J.~C. Cuillandre, T. Forveille, P. Martin, N. Manset, J. Shapiro)
and director (G. Fahlman) as well as the Terapix MEGACAM data
reduction center staff (Y. Mellier, G. Missonier). 
Research at PSI has been financially supported by the Swiss National Science 
Foundation (grant 20-66875.01). We thank the International Space Science 
Institute (ISSI) in Bern for further significant financial support to the 
{\it XMM-Newton} team. {\it XMM-Newton}  is an ESA science mission with 
instruments and  contributions directly funded by ESA Member States 
and the USA (NASA). 

\bigskip

\centerline\textbf{ REFERENCES}
\bigskip
 \parskip=0pt
 {\small
 \baselineskip=11pt
 \refs Adams F.~C., Lada C.~J., and Shu F.~H. (1987) {\em Astrophys. J., 312}, 788-806.
 \refs d'Alessio P., Calvet N., Hartmann  L., Lizano S., and Cant\'o  J. (1999) {\em Astrophys. J., 527}, 893-909.
 \refs Andrews S.~M. and  Williams J.~P. (2005) {\em Astrophys. J., 631}, 1134-1160. 
 \refs Anglada G. (1995) {\em Rev. Mex. Astron. Astrophys., 1}, 67-76.   
 \refs Apai D., Pascucci I., Henning Th., Sterzik M.~F., Klein R., Semenov D., 
       G\"unther E., and Stecklum B. (2002) {\em Astrophys. J., 573}, L115-117.
 \refs Argiroffi C., Drake J.~J., Maggio A., Peres G., Sciortino S.,  and  Harnden F.~R.
	(2004) {\em Astrophys. J., 609}, 925-934.
 \refs Balbus S.~A. and Hawley J.~F. (1991) {\em Astrophys. J., 376}, 214-233.  
 \refs Bally J.,  Feigelson E.,  and   Reipurth B. (2003) {\em Astrophys. J., 584}, 843-852.   
 \refs Barrado y Navascu\'es D.  and  Mart\'{\i}n E. (2003) {\em Astron. J., 126}, 2997-3006.
 \refs Beichman C.~A., Myers P.~C., Emerson J.~P., Harris S., Mathieu R., Benson P.~J.,
     and  Jennings R.~E. (1986) {\em Astrophys. J., 307}, 337-349.
 \refs Bertin, E.  and  Arnouts S. (1996), {\em Astron. Astrophys. Suppl., 117}, 393-404.
 \refs Boulade O., et al. (2003) In {\em Instrument Design and Performance 
	 for Optical/Infrared Ground-based Telescopes} (M. Iye  and  A.~F.~M.  Moorwood, eds.), pp. 72-81.
	 The International Society for Optical Engineering.
 \refs Bouvier J. (1990) {\em Astron. J., 99}, 946-964.
 \refs Brice\~no C., Hartmann L., Stauffer J.,  and   Mart{\'{\i}}n E. (1998), {\em Astron. J., 115},
	2074-2091.
 \refs Brice\~no C., Luhman K.~L., Hartmann L., Stauffer J.~R.,  and  Kirkpatrick J.~D.
	  (2002) {\em Astrophys. J., 580}, 317-335. 
 \refs Burrows A. Sudarsky D.,  and   Lunine J.~I. (2003) {\em Astrophys. J., 596}, 587-596.
 \refs Chabrier G. (2003)  {\em Publ. Astron. Soc. Pac., 115}, 763-795.
 \refs Chabrier G., Baraffe I., Allard F.,  and  Hauschildt P. (2000) {\em Astrophys. J., 542}, 464-472.
 \refs Cuillandre J., Luppino G.~A., Starr B.~M.,  and	Isani S. (2000) {\em SPIE, 4008}, 1010-1021.
 \refs Damiani F.  and  Micela G. (1995) {\em Astrophys. J., 446}, 341-349.
 \refs Damiani F., Micela G., Sciortino S.,  and  Harnden F.~R. Jr.
	  (1995) {\em Astrophys. J., 446}, 331-340.
 \refs  Dobashi K., Uehara H., Kandori R., Sakurai T., Kaiden M.,
	    Umemoto T., and Sato, F. (2005) {\em Publ. Astron. Soc. Japan, 57}, S1-S368. 
 \refs Duch\^ene G., Monin J.-L., Bouvier J.,  and  M\'enard F. (1999) {\em Astron. Astrophys., 351}, 954-962.
 \refs Durney B.~R., De Young D.~S., and Roxburgh, I.~W. (1993) {\em Solar Phys. 145}, 207-225. 
 \refs Eisl\"offel J. and  Mundt R. (1998) {\em Astron. J., 115}, 1554-1575. 
 \refs Eisner J.~A., Hillenbrand L.~A., Carpenter J.~M., and  Wolf S. (2005) {\em Astrophys. J., 635}, 396-421. 
 \refs Evans N.~J., et al. (2003) {\em Publ. Astron. Soc. Pac., 115}, 965-980.
 \refs Feigelson E.~D.  and  Montmerle T. (1999) {\em Ann. Rev. Astron. Astrophys., 37}, 363-408. 
 \refs Feigelson E.~D., Jackson J.~M., Mathieu R.~D., Myers P.~C.,
	   and  Walter F.~M. (1987) {\em Astron. J., 94}, 1251-1259.
 \refs Flaccomio E., Micela G., Sciortino S., Damiani F., Favata F., Harnden F.~R. Jr.,  and  Schachter J.
	  (2000) {\em Astron. Astrophys., 355}, 651-667.
 \refs Flaccomio E., Micela G.,  and  Sciortino S.  (2003a) {\em Astron. Astrophys., 402}, 277-292.
 \refs Flaccomio E., Damiani F., Micela G., Sciortino
	  S., Harnden F.~R. Jr., Murray  S.~S.,  and  Wolk S.~J. (2003b) {\em Astrophys. J., 582}, 398-409.
 \refs Fleming T.~A., Giampapa M.~S., Schmitt J.~H.~M.~M.,  and  Bookbinder J.~A. (1993)
       {\em Astrophys. J., 410}, 387-392.
 \refs Galfalk M., et al. (2004) {\em Astron. Astrophys., 420}, 945-955.     
 \refs Garcia-Alvarez D., Drake J.~J., Lin L., Kashyap V.~L.,  and   Ball B. (2005)
	{\em Astrophys. J., 621}, 1009-1022.
 \refs Ghez A.~M., Neugebauer G., and Matthews K. (1993) {\em Astron. J., 106}, 2005-2023.
 \refs Glassgold A.~E.,  Najita J.,  and  Igea J. (2004) {\em Astrophys. J., 615}, 972-990.  
 \refs G\'omez M., Hartmann L., Kenyon S.~J., and Hewett R. (1993) {\em Astron. J., 105}, 1927-1937.
 \refs Goodwin S.~P.,  Whitworth A.~P.,  and  Ward-Thompson D. (2004) {\em Astron. Astrophys., 419}, 543-547. 
 \refs Grosso N., et al. (2006a) {\em Astron. Astrophys.}, in press.
 \refs Grosso N., Audard M., Bouvier J., Briggs K., and G\"udel M. (2006b) {\em Astron. Astrophys.}, in press.
 \refs G\"udel M. (2004) {\em Astron. Astrophys. Rev., 12}, 71-237. 
 \refs G\"udel M. (2006a) {\em Astron. Astrophys.}, in press. 
 \refs G\"udel M. (2006b) {\em Astron. Astrophys.}, in press. 
 \refs G\"udel M., Guinan E.~D.,  and  Skinner S.~L. (1997) {\em Astrophys. J., 483}, 947-960.
 \refs G\"udel M., Skinner, S.~L., Briggs, K.~R., Audard, M., Arzner, K., and Telleschi, A. (2005), 
	 {\em Astrophys. J., 626}, L53-56. 
 \refs Guieu S., Pinte C., Monin J.-L., M\'enard F., Fukagawa M., Padgett D., Carey S., Noriega-Crespo A.,
       and  Rebull L. (2005) In {\it PPV Poster Proceedings} \\
       http://www.lpi.usra.edu/meetings/ppv2005/pdf/8096.pdf
 \refs Guieu S., Dougados C., Monin J.-L., Magnier E., and  Mart\'{\i}n E.~L. (2006) {\em Astron. Astrophys., 446}, 
	 485-500.  
 \refs Gullbring E., Hartmann L., Brice\~no, C., and Calvet, N. (1998) {\em Astrophys. J., 492}, 323-341
 \refs Hartmann L. Megeath S.~T., Allen L., Luhman K., Calvet N., D'Alessio P., 
       Franco-Hernandez R., and Fazio G. (2005) {\em Astrophys. J., 629}, 881-896.
 \refs Hartigan P.  and  Kenyon S.~J. (2003) {\em Astrophys. J., 583}, 334-357. 
 \refs Hirth G.~A., Mundt R.,  and  Solf J. (1997) {\em Astron. Astrophys. Suppl., 126}, 437-469.   
 \refs Jayawardhana R., Mohanty S., and Basri G. (2002) {\em Astrophys. J., 578}, L141-144.
 \refs Jayawardhana R., Mohanty S., and Basri G. (2003) {\em Astrophys. J., 592}, 282-287.
 \refs Kaifu, N., et al. (2004) {\em Publ. Astron. Soc. Japan, 56}, 69--173.
 \refs Kastner J.~H., Huenemoerder D.~P.,
	  Schulz N.~S., Canizares C.~R.,  and	Weintraub D.~A. (2002) {\em Astrophys. J., 567}, 434-440.
 \refs Kenyon S.~J. and Hartmann L. (1995) {\em Astrophys. J. Suppl., 101}, 117-171. 
 \refs Kenyon S.~J., Hartmann L.~W., Strom K.~M., and  Strom S.~E. (1990) {\em Astrophys. J., 99}, 869-887.
 \refs Kenyon S.~J., Dobrzycka D., and Hartmann L. (1994) {\it Astron. J., 108}, 1872-1880.
 \refs Kroupa P.  and  Bouvier J. (2003) {\em Mon. Not. R. Astron. Soc., 346}, 343-353. 
 \refs Kun M. (1998) {\em Astrophys. J. Suppl., 115}, 59-89.
 \refs Leinert Ch., Zinnecker H., Weitzel N., Christou J., Ridgway S.~T., Jameson R., Haas M.,  and  
       Lenzen R. (1993) {\it Astron. Astrophys., 278}, 129-149.
 \refs Liu M.~C., Najita J., and  Tokunaga A.~T. (2003) {\em Astrophys. J., 585}, 372-391.
 \refs Loinard L., Mioduszewski A.~J., Rodr\'{\i}guez L.~F., Gonz\' alez R.~A., 
       Rodr\'{\i}guez M.~I., and Torres R.~M. (2005) {\it Astrophys. J., 619}, L179-L182.
 \refs Luhman K.~L. (2000) {\em Astrophys. J., 544}, 1044-1055.
 \refs Luhman K.~L. (2004) {\em Astrophys. J. 617}, 1216-1232.
 \refs Luhman K.~L., Brice\~no C., Stauffer J.~R.,  Hartmann L., Barrado Y Navascu\'es, D.,  and 
       Caldwell N.  (2003) {\em Astrophys. J. 590}, 348-356.
 \refs Magnier E.~A.  and Cuillandre J.-C. (2004) {\em Publ. Astron. Soc. Pac., 116}, 449-464.
 \refs Mart{\'{\i}}n E.~L.,  Dougados C., Magnier E., M{\' e}nard F.,  Magazz{\`u} A., 
       Cuillandre J.-C.,  and	Delfosse X., (2001) {\em Astrophys. J., 561}, L195-198.
 \refs Mathieu R.~D. (1994) {\em Ann. Rev. Astron. Astrophys., 32}, 465-530.
 \refs Mohanty S., Jayawardhana R., and  Basri G. (2005)  {\em Astrophys. J., 626}, 498-522.
 \refs Monin J.-L., Dougados C., and Guieu S. (2005) {\em Astron. Nachrichten, 326}, 996-1000. 
 \refs Montmerle T., Grosso N., Tsuboi Y., and Koyama K. (2000) {\em Astrophys. J., 532}, 1097-1110.
 \refs Moraux E., Bouvier J., Stauffer J.~R., and Cuillandre J.-C. (2003) {\em Astron. Astrophys., 400}, 891-902.
 \refs Muzerolle J., Luhman K.~L., Brice\~no C., Hartmann L., and Calvet N. (2005)  {\em Astrophys. J., 625},
	 906-912.
 \refs Myers P.~C., Fuller G.~A., Mathieu R.~D., Beichman C.~A., Benson P.~J., Schild R.~E., and
       Emerson J.~P. (1987) {\em Astrophys. J., 319}, 340-357.
 \refs Neuh\"auser R., Sterzik M.~F., Schmitt J.~H.~M.~M.,
	  Wichmann R.,  and  Krautter J. (1995) {\em Astron. Astrophys., 297}, 391-417.
 \refs Noyes R.~W., Hartmann, L.~W., Baliunas S.~L., Duncan D.~K., and  Vaughan  A.~H. (1984) {\it Astrophys. J., 279}, 763-777.
 \refs Onishi T., Mizuno A., Kawamura A., Tachihara K.,  and  Fukui Y. (2002), {\em Astrophys. J., 575}, 950-973.
 \refs Padgett D.~L., et al.  (2004) {\em Astrophys. J. Suppl., 154}, 433-438. 
 \refs Papovich C., et al. (2004) {\em Astrophys. J. Suppl., 154}, 70-74. 
 \refs Pizzolato N., Maggio A., Micela G., Sciortino S., and Ventura P. (2003) {\em Astrophys. J., 397}, 147-157.
 \refs Pravdo S.~H., Feigelson E.~D., Garmire G., Maeda Y., Tsuboi Y.,  and  Bally J. (2001) 
	{\em Nature, 413}, 708-711.
 \refs Preibisch T.  and  Zinnecker H. (2001) {\em Astron. J., 122}, 866-875.
 \refs Preibisch T., et al. (2005a) {\em Astrophys. J. Suppl., 160}, 401-422.
 \refs Preibisch T., et al. (2005b) {\em Astrophys. J. Suppl., 160}, 582-593.
 \refs Raga A.~C., Noriega-Crespo A.,  and  Vel\' azquez P.~F. (2002), {\em Astrophys. J., 576}, L149-152.
 \refs Reipurth B.  and  Clarke C. (2001) {\em Astron. J., 122}, 432-439.
 \refs Rieke G.~H., et al. (2004) {\em Astrophys. J. Suppl., 154}, 25-29. 
 \refs Scelsi L., Maggio, A., Peres G.,  and	Pallavicini R. (2005) {\em Astron. Astrophys., 432}, 671-685.
 \refs Siess L., Dufour E.,  and  Forestini M. (2000) {\em Astron. Astrophys., 358}, 593-599.
 \refs Simon M., et al. (1995) {\em Astrophys. J., 443}, 625-637.
 \refs Slesnick C.~L., Hillenbrand L.~A.,  and   Carpenter J.~M. (2004), {\em Astrophys. J., 610}, 1045-1063.
 \refs Stapelfeldt K.~R. and Moneti A. (1999) In {\em The Universe as Seen by ISO} (P. Cox  and  
	 M.~F. Kessler, eds.), pp. 521-524. European Space Agency.
 \refs Stelzer B. and  Neuh\"auser R. (2001)  {\em Astron. Astrophys., 377}, 538-556. 
 \refs Strom K.~M.  and  Strom S.~E. (1994) {\em Astrophys. J., 424}, 237-256.
 \refs Strom K.~M.,  Strom S.~E., Edwards S., Cabrit S.,  and  Skrutskie M. (1989) {\em Astron. J., 97}, 1451-1470.
 \refs Strom K.~M., Strom S.~E., Wilkin F.~P., 
	  Carrasco L., Cruz-Gonzalez I., Recillas E., Serrano A., Seaman R.~L., Stauffer J.~R., Dai D., 
	   and  Sottile J. (1990) {\em Astrophys. J., 362}, 168-190.
 \refs Telleschi A., G\"udel M.,  Briggs K., Audard M., Ness J.-U.,  and  Skinner S.~L. (2005)
       {\em Astrophys. J., 622}, 653-679.
 \refs Walter F.~M., Brown A., Mathieu R.~D., Myers P.~C.,  and 
	  Vrba F.~J. (1988) {\em Astron. J., 96}, 297-325.
 \refs Weaver W.~B. and Jones G. (1992) {\em Astrophys. J. Suppl., 78}, 239-266.
 \refs White R.~J. and Basri, G. (2003) {\em Astrophys. J., 582}, 1109-1122.
 \refs White R.~J. and Ghez A.~M. (2001) {\em Astrophys. J., 556}, 265-295.  
 \refs White R.~J. and Hillenbrand L.~A. (2004) {\em Astrophys. J., 616}, 998-1032.
 \refs Young K.~E.,  et al. (2005) {\em Astrophys. J., 628}, 283-297.

\end{document}